\documentclass[journal=jacsat,manuscript=article]{achemso}

\usepackage[version=3]{mhchem} 
\usepackage{mciteplus}
\usepackage{makecell}
\usepackage{xspace}
\usepackage{textcomp}

\author{Peter M. Piechulla}
\affiliation[TU Delft]
{Department of Chemical Engineering, Delft University of Technology, 2629 HZ, Delft, The Netherlands}
\email{p.m.piechulla@tudelft.nl}
\author{Iris Blommestein}
\affiliation[TU Delft]
{Department of Chemical Engineering, Delft University of Technology, 2629 HZ, Delft, The Netherlands}
\author{Anne van Gerven}
\affiliation[TU Delft]
{Department of Chemical Engineering, Delft University of Technology, 2629 HZ, Delft, The Netherlands}
\author{Henri Pelzer}
\affiliation[TU Delft]
{Department of Chemical Engineering, Delft University of Technology, 2629 HZ, Delft, The Netherlands}
\author{Ezgi Demiröz}
\affiliation[TU Delft]
{Department of Chemical Engineering, Delft University of Technology, 2629 HZ, Delft, The Netherlands}
\author{Jonas Englhard}
\affiliation[FAU]{Chair ’Chemistry of Thin Film Materials’, Materials Chemistry section, Department of Chemistry and Pharmacy, Friedrich-Alexander-Universität Erlangen-Nürnberg,  IZNF, Cauerstr. 3, 91058 Erlangen, Germany}
\author{Mingliang Chen}
\affiliation[TU Delft]
{Department of Chemical Engineering, Delft University of Technology, 2629 HZ, Delft, The Netherlands}
\author{Julien Bachmann}
\affiliation[FAU]{Chair ’Chemistry of Thin Film Materials’, Materials Chemistry section, Department of Chemistry and Pharmacy, Friedrich-Alexander-Universität Erlangen-Nürnberg,  IZNF, Cauerstr. 3, 91058 Erlangen, Germany}
\author{J. Ruud van Ommen}
\affiliation[TU Delft]
{Department of Chemical Engineering, Delft University of Technology, 2629 HZ, Delft, The Netherlands}

\title[Direct ALD anodes]{Ionomer-free Iridium Anodes with Low Loading for Zero-gap Electrolyzers via Direct Atomic Layer Deposition}


\begin{document}

\newcommand{\Acm}{\,A$\cdot$cm$^{-2}$\xspace}
\newcommand{\mAcm}{\,mA$\cdot$cm$^{-2}$\xspace}
\newcommand{\mVdec}{\,mV$\cdot$dec$^{-1}$\xspace}
\newcommand{\mOcm}{\,m$\Omega\cdot$cm$^{-2}$\xspace}
\newcommand{\atp}{\,\%$_{\text{at}}$\xspace}
\newcommand{\uVh}{\,µV$\cdot$h$^{-1}$\xspace}
\newcommand{\ugcm}{\,µg$\cdot$cm$^{-2}$\xspace}
\newcommand{\mgcm}{\,mg$\cdot$cm$^{-2}$\xspace}
\newcommand{\um}{\,µm\xspace}
\newcommand{\nm}{\,µm\xspace}
\newcommand{\kWg}{\,kW$\cdot$g$_{\text{Ir}}$\xspace}
\newcommand{\gkW}{\,g$_{\text{Ir}}\cdot$kW$^{-1}$\xspace}

\begin{tocentry}

\includegraphics[]{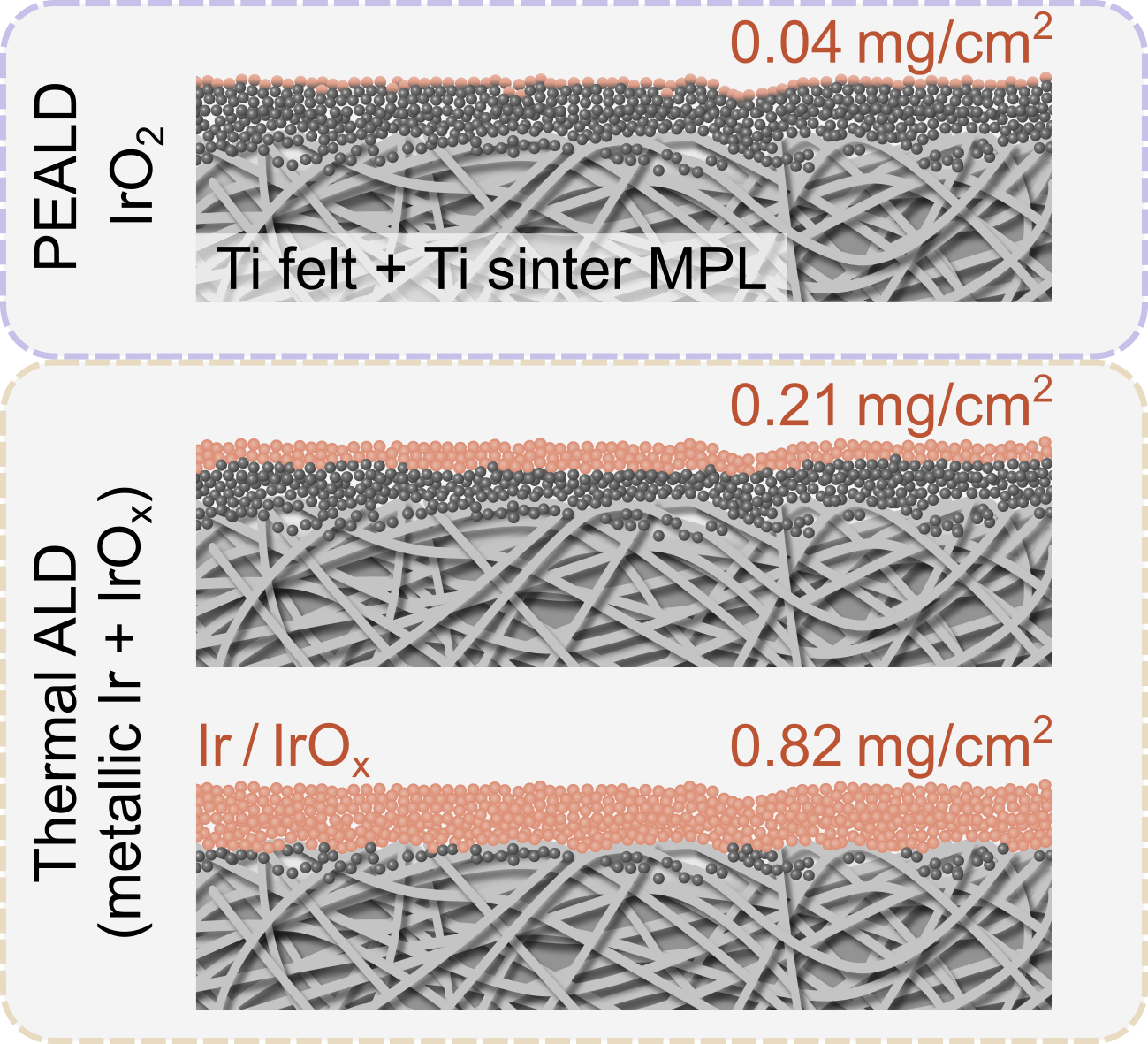}

Iridium-based anode catalyst layers for proton exchange membrane water electrolysis were fabricated via ALD. Using custom-made microporous substrates and thermal ALD, current densities of up to 2.5\Acm at 0.21\,mg/cm$^2$ Ir loading were achieved. By leveraging the localized Ir deposition of PEALD at the membrane-catalyst layer interface, even lower loadings of down to 0.04\,mg/cm$^2$ were accessible.

\end{tocentry}

\begin{abstract}
Proton exchange membrane water electrolysis (PEMWE) is among the most promising techniques to provide hydrogen as both energy carrier and feedstock for the chemical industry, due to its compatibility with intermittent renewable electricity sources.
However, a major bottleneck in scaling up PEMWE is the limited supply of the anode catalyst Ir.
This work shows a novel approach to tailoring ionomer-free catalyst layers based on custom-made microporous substrates and two different atomic layer deposition (ALD) processes, both using Ir(EtCp)(CHD) precursor.
Thermal ALD using \ce{O3} as counter-reactant yields mostly metallic Ir/IrO$_x$ and an electrode Ir loading of 0.21\mgcm. The PEMWE cell-level performance is comparable to a commercial catalyst-coated membrane anode despite $7\times$ lower loading, and is maintained over an accelerated durability test equivalent to 600\,h of intermittent operation.
Moreover, using a low-loading ALD-made cathode, the total platinum group metal loading is only 0.24\mgcm.
Plasma-enhanced ALD reduces the Ir loading to 0.04\mgcm by tailoring the catalyst distribution, however, yields lower-activity \ce{IrO2}.
Finally, direct ALD catalyst deposition needs no ionomer-bearing ink. 
The anodes are therefore compatible with various membrane materials, as demonstrated in a \ce{CO2} electrolysis experiment with an anion exchange membrane.

\end{abstract}


\section{Introduction}

The energy transition is in full swing, and solar and wind are now among the least expensive sources of newly installed electrical power.
The new frontier is electrification of material cycles which goes hand in hand with medium-term energy storage technology using energy carriers such as hydrogen.
Zero-gap, membrane-based electrolysis is among the most promising technologies addressing both these challenges because of its high power densities and compatibility with intermittent electricity supply.
Proton exchange water membrane water electrolysis (PEMWE) is a prominent example, providing hydrogen both as a feedstock for the production of base chemicals and as an energy carrier, thereby replacing fossil fuels in both of these functions. 
Alternatively, membrane-based electrolysis exhibits significant potential to produce higher-value chemicals directly, such as direct electrolysis of \ce{CO2} to hydrocarbons.
Key components for any electrolyzer are effective and stable electrocatalysts for both the anode and cathode half-reactions.

Platinum group metal (PGM) catalysts are highly active and exceptionally stable, but are also scarce and expensive.
Especially for Ir as the classic catalyst for the oxygen evolution reaction (OER), inelasticity of the supply may soon pose a major bottleneck when scaling electrolysis technology to the magnitude required to replace fossil fuels.
Technological roadmaps\cite{Wang2025,GreenHydrogenCost2020} for PEMWE therefore include goals to reduce the required amount of Ir from typically $>1$\mgcm, for example to $\leq$0.2\mgcm or, in terms of power-specific loading $\leq$0.4 g$\cdot$kW$^{-1}$.

Traditional wet deposition methods for catalyst layer fabrication reach limits in reducing the loading in the sub--\mgcm range.\cite{Bernt2018,Padgett2023}
The approach consists of multiple steps: first, catalyst particles are synthesized; second, the particles are combined with a polymer electrolyte binder and further additives into an ink or slurry formulation; third, the ink/slurry is deposited onto a substrate by means of spray coating, doctor blading or slot-die coating.
Especially in thin catalyst layers, the homogeneity of catalyst distribution is ultimately limited by the finite size of the catalyst particles, which leads to non-contiguous layers and thus poor electrical contact between catalyst particles and porous transport layers (PTLs).\cite{Bernt2018}
Though low power-specific Ir loadings (high mass-activity) can be achieved, the inhomogeneity harms the performance of the electrolyzer by increasing the overpotential.

To simultaneously achieve both low power-specific Ir loading and low overpotential, methods that directly deposit catalytically active material onto a PTL have recently attracted attention as an alternative to traditional wet deposition.\cite{ChenReview2025}
This approach reduces the complexity of fabrication by uniting the three steps of catalyst synthesis, ink/slurry formulation and deposition into a single deposition step.
The resulting porous transport electrodes (PTEs) are ionomer-free, and are therefore compatible with a variety of membrane materials besides the traditional acidic, fluorinated ionomers such as Nafion\texttrademark.
Ionomer-free PTEs can be made through a variety of methods, ranging from wet-chemical\cite{Xie2024} and electrodeposition\cite{Ding2024}, to physical vapor deposition (sputtering)\cite{Lee2023}. 
Through all of these methods, low Ir weight loadings between 0.08 and 0.34\mgcm have been achieved, while simultaneously reaching similar or higher currents than their conventional CCM references, respectively. 
Notably, all of the cited studies included pretreatment steps to increase the surface area before Ir deposition\cite{Lee2023}, or used proprietary substrates and the Ir catalyst itself comprised a nano structure.\cite{Ding2024,Xie2024}.

Atomic layer deposition (ALD) is well known for its facility to conformally coat rough and three-dimensional structures\cite{2021vanOmmen}, its wide range of attainable catalyst materials, and facility to tailor catalyst morphologies on the nano-scale\cite{PiechullaALDpm2026}. 
Using ALD to coat commercial carbon-based microporous layers (MPLs) directly with Pt, we recently demonstrated PEMWE PTE cathodes with exceptionally low Pt loadings in the range of 1 to 5\ugcm, which is about 100$\times$ lower than the benchmark commercial CCM cathode\cite{Chen2026}. These cathode PTEs achieved similarly low overpotentials and maintained a stable performance over an accelerated durability test (ADT).
This approach, in particular in combination with the recent advances in scalable, spatial ALD technology bears significant potential to reduce both Pt consumption and complexity of catalyst layer fabrication.\cite{vandePol2025,Diaz2024}
Similarly, ALD has been explored to fabricate OER electrodes based on Ir,\cite{Schlicht2019} but such anodes in a PEMWE experiment have thus far been limited to current densities $<$0.2\Acm\cite{Laube2021}.
Due to the sluggish OER kinetics, high currents at acceptable overpotentials are difficult to achieve.\cite{McCrory2015,Matsumoto1986}
An obvious solution is compensating the lower activity by increasing the catalytically active surface\cite{vandePol2025}.
However, commercially available PTLs with MPLs that provide high specific surface areas to deposit Ir on are typically carbon-based\cite{Chen2026}, and therefore not suitable for the oxidative potentials at the anode.
Additionally, it is unclear how the particular growth behavior of ALD will affect the catalyst activity and loading of a PTE, since ALD is known for its conformality whereby the coating can reach deep into porous structures.\cite{2021vanOmmen}

\begin{figure}
    \centering
    \includegraphics[width=0.8\linewidth]{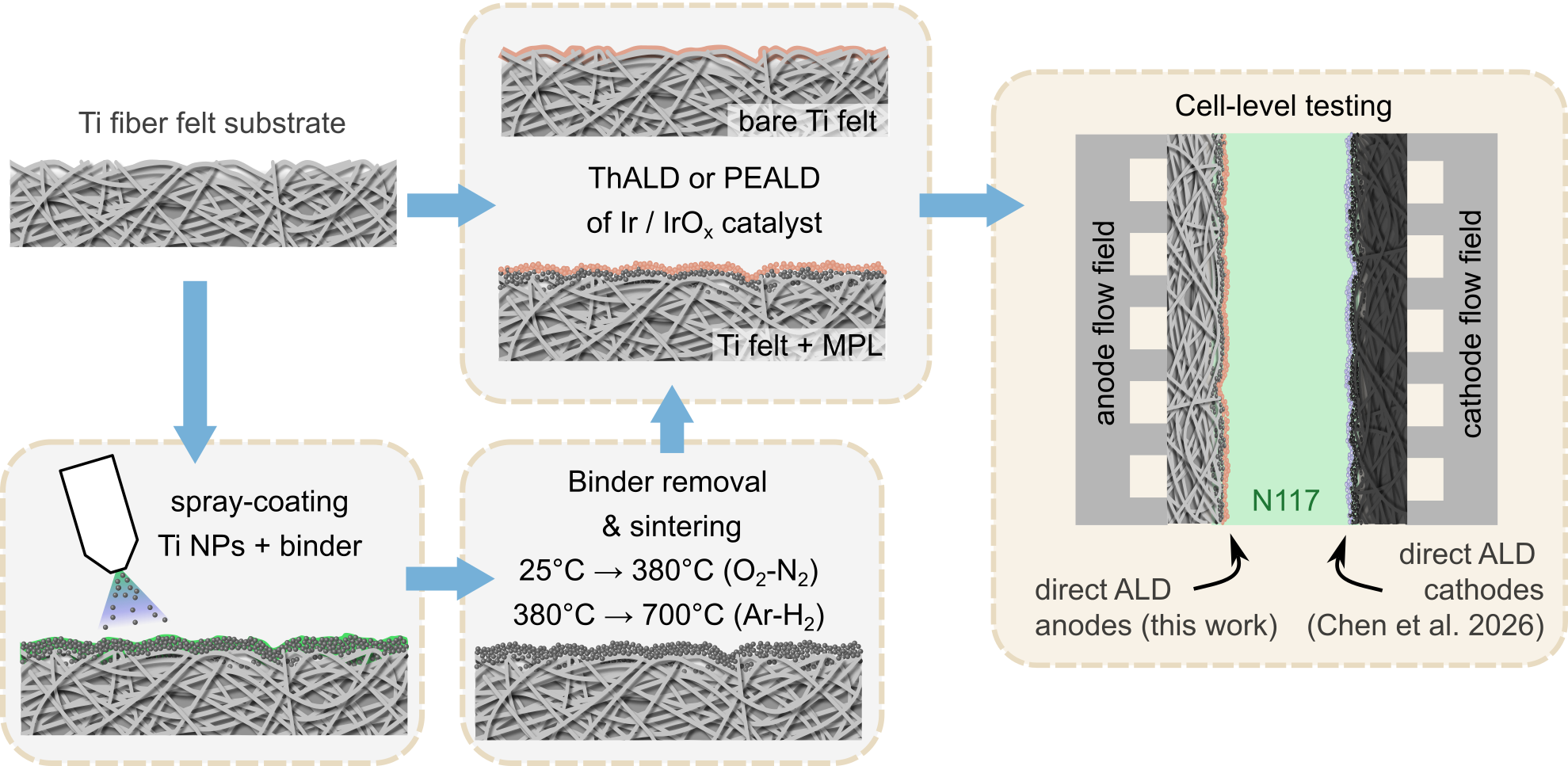}
    \caption{Schematic illustration of the PTE fabrication approach demonstrated in this work. Ti fiber felts are either directly coated with Ir via ALD, or first equipped with an MPL, and then coated with Ir. The fabricated anodes are tested against ALD-made ionomer-free cathodes presented by Chen et al.\cite{Chen2026}}
    \label{fig:overview}
\end{figure}

In this study, we investigate the interplay of direct ALD of Ir/IrO$_x$ and custom-made MPLs to fabricate low Ir loading anodes for PEMWE (Figure~\ref{fig:overview}).
We begin by showing the differences between oxidization state and activity of PEMWE anode catalysts made by thermal ALD (ThALD) and plasma-enhanced ALD (PEALD).
We then enhance the surface area and activity of the respective catalysts by introducing an MPL made from sintered Ti nanoparticles.
Crucially, the catalyst distribution within the MPL is accessible through the choice of deposition process (ThALD or PEALD), which affects the Ir loading and performance of the respective anodes. 
ThALD MPL anode shows stable performance, whereas a PEALD anode shows significant degradation.
For the case of a ThALD MPL anode, we discuss the individual overpotential contributions (charge transfer, ohmic, mass transport) which we distinguish through electrochemical impedance spectroscopy (EIS).
This analysis reveals a qualitative difference in mass transport overpotential compared to a traditional, wet-deposited CCM anode.
Zooming out from the discussion of individual anodes, we discuss the power-specific Ir loading together with the current density for a range of different ALD-made anodes, and show that each configuration of ALD process and PTL/MPL substrates requires its own loading optimization.
Finally, we demonstrate the versatility of ALD-made ionomer-free anodes by showing their function in an anion exchange membrane (AEM)-based \ce{CO2} electrolyzer.

\section{Results}

\subsection{ALD of iridium}

Two different ALD processes, ThALD and PEALD, are investigated throughout this study.
Both processes used the same Ir(EtCp)(CHD) precursor, and were carried out at 220\,°C reactor temperature. 
Further parameters are described in the Methods section, and results are discussed in the following.

\subsubsection{Thermal ALD}
The ThALD process has been presented in previous studies\cite{Schlicht2018,Schlicht2018a}, and uses \ce{O3} as a counter-reactant.
To study the chemical composition and oxidation state, we analyzed a layer deposited by 75 ThALD cycles on planar reference samples using X-ray photoelectron spectroscopy (XPS).
A survey scan (Supporting Information Figure~S1a) of a ThALD sample showed a composition of 48.9\atp~Ir, 36.4\atp~O and 14.9\atp~C; the latter could stem from unreacted precursor residues or adventitious C from the environment.
The underlying Ti of the substrate was not detected which indicates that 75 cycles led to a closed film. 

\begin{figure}
    \centering
    \includegraphics[width=0.75\linewidth]{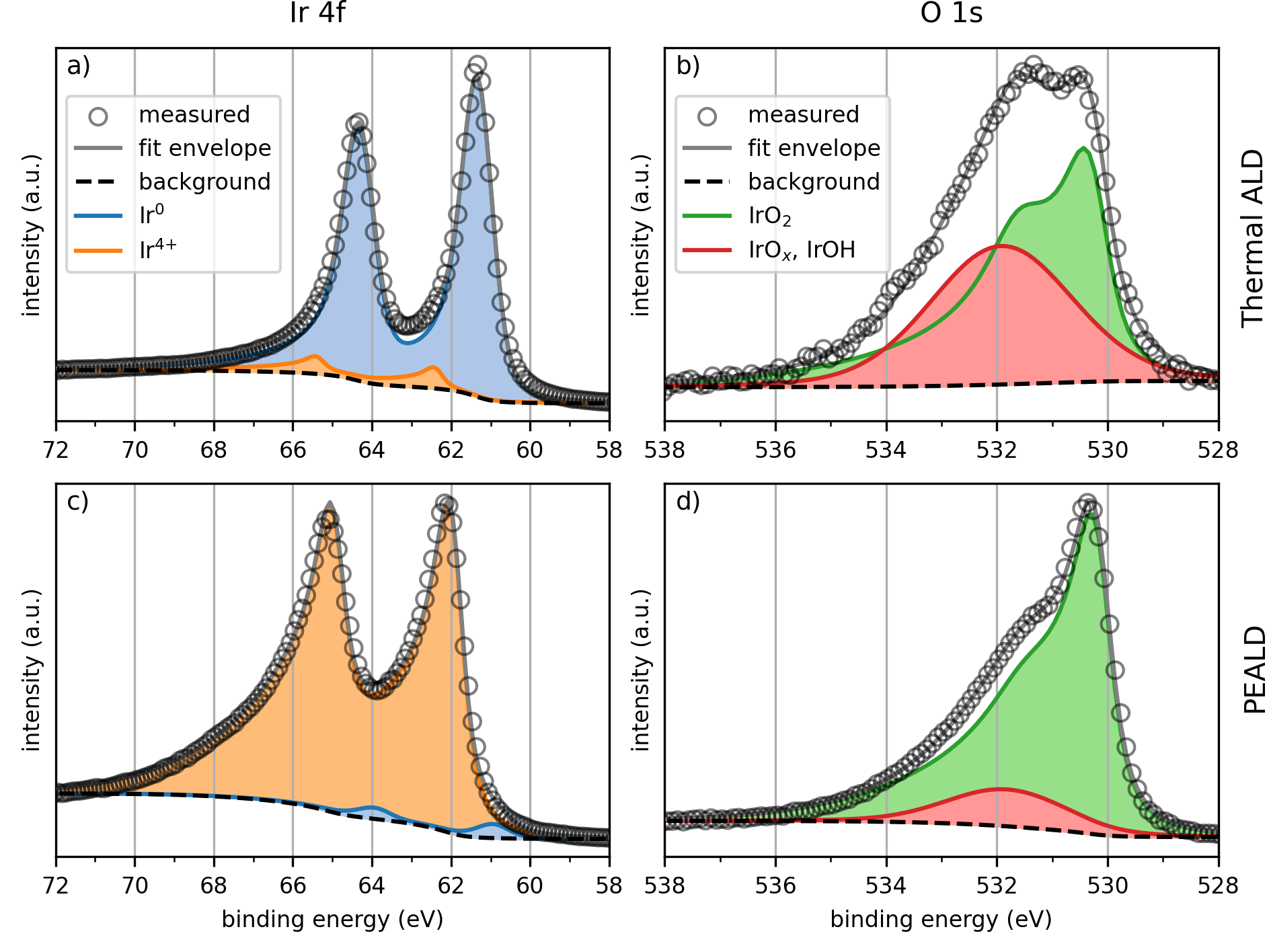}
    \caption{High resolution XPS scans of the Ir 4f (a, c) and O 1s (b, d) region for ThALD (a,b) and PEALD (c,d).}
    \label{fig:XPS}
\end{figure}

The low ratio [O]/[Ir] suggests that Ir is to a large part metallic. 
Fitting to the Ir 4f peak region of a high-resolution XPS scan (Figure~\ref{fig:XPS}a) indeed confirms that the ThALD process produces metallic Ir$^0$ (88\,\%), which contained only about 12\,\% Ir$^\text{4+}$.
To elucidate the nature of the Ir oxide detected, we further analyzed the O 1s region (Figure~\ref{fig:XPS}b).
The peak near 530\,eV and a satellite peak (summarized in green in Figure~\ref{fig:XPS}b) is associated with \ce{IrO2}\cite{Geiger2018,Freakley2017}.
The higher binding energy peak (around 532\,eV, red) is associated with IrOH\cite{Freakley2017} or other IrO$_x$ types.\cite{Geiger2018}.
We conclude that the metallic Ir layer is terminated with Ir oxide, which may be partly hydroxylated.

According to X-ray reflectometry (XRR) measurements (Supporting Information Figure~S2), the film was approximately 1.2\,nm thick and the density lower than expected for metallic Ir, indicating a rough layer.
However, we note that the layer was close to the lower detection limit of the XRR method, therefore these values are rather indicative, i.e., could vary by $\pm$ 1 nm.
Upon increasing the number of cycles to 150, the layer thickness increased to 5\,nm, and the density approached that of a dense metallic Ir layer.
These two experiments indicate a growth per cycle (GPC) $\approx 0.5$\,Å, which is lower than the 0.6--0.9\,Å reported earlier for the same process\cite{Schlicht2018a,Haschke2019}, however, these earlier studies used different substrates.
Also, we observe a significant growth delay of about 50 cycles due to due to substrate-inhibited nucleation (Supporting Information Figure~S4), which is also observed for the Ir loading of PTL substrates that will be discussed later (Supporting Information Figure~S7).

\subsubsection{Plasma-enhanced ALD}
To explore the potential of different ALD processes, we additionally developed a PEALD processes that uses \ce{O2} plasma as a counter-reactant.
The XPS survey scan of a PEALD sample (Supporting Information Figure~S1b) showed a composition of 27.5\atp~Ir, 61.7\atp~O and 10.8\atp~C.
The ratio [O]/[Ir]=2.2$\approx$2 indicates the presence of \ce{IrO2}, which is supported by the Ir 4f region fit showing 97\,\%  Ir$^\text{4+}$ (Figure~\ref{fig:XPS}c) and $<3$\,\% Ir$^0$.
The O 1s region further shows the distinct peak shape of crystalline \ce{IrO2}\cite{Reier2014,Geiger2018}. 
Fitting different components to the O 1s region indeed supports the hypothesis that the layer consisted of \ce{IrO2}\cite{Geiger2018,Freakley2017}, with a minority of IrO$_x$ or IrOH that likely terminates the surface.
Notably, this is different from recent results of van de Poll et al., who observed mostly metallic Ir from PEALD at reactor temperatures $T>200$\,°C.\cite{vandePol2025} However, they used \ce{N2} as carrier gas at atmospheric pressure instead of the Ar carrier gas and vacuum conditions used in the PEALD of the present study.

Judging from the XRR-determined layer thicknesses (Supporting Information Figure~S3), \ce{IrO2} growth shows both a delay at low number of cycles, and a decrease of GPC beyond 300 cycle (Supporting Information Figure~S4). On average, a GPC\,$\approx 0.3$\,Å was obtained, which is slightly lower than $\approx 0.4$\,Å as previously reported for PEALD processes using the same precursor.\cite{vandePol2025}
Alongside \ce{IrO2} film growth, we observe a growth of the underlying TiO$_x$ layer (Supporting Information Figure~S3) to up to 11\,nm thickness after 600 PEALD cycles. 

\subsection{OER activities of bare Ir layers}

\begin{figure}
    \centering
    \includegraphics[width=0.95\linewidth]{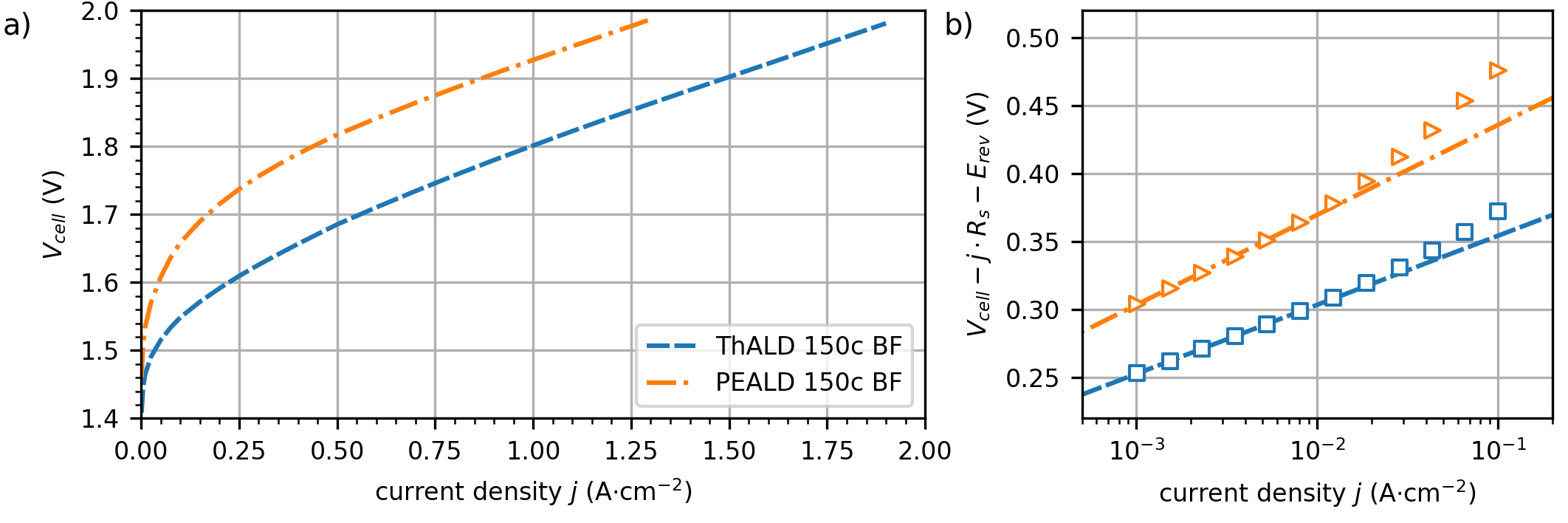}
    \caption{Performance of porous transport electrodes (PTEs) made by ALD of Ir/\ce{IrO2} onto bare Ti felt (BF) PTLs. a) Polarization curves. b) Tafel plots with fitted slopes in up to $10^{-2}$\Acm.}
    \label{fig:ThvsPE}
\end{figure}

To compare the OER activities of catalyst materials from the ThALD and PEALD process, we coated two bare, commercial Ti fiber felt PTLs (Figure~\ref{fig:SEM}a) with 150 cycles of either one process (Figure~\ref{fig:ThvsPE}).
We tested these anode PTEs in a PEMWE single cell experiment using ALD-made Pt PTEs as cathodes for which we demonstrated high and stable performance despite a Pt loading of only 22\ugcm in a previous work\cite{Chen2026}.
A direct comparison between a commercial, double-sided CCM and the anode CCM used as reference in this work is shown in Supporting Information Figure~S5, indicating no limitation from the PTE cathode used here.

Both anodes achieved currents $>1$\Acm~within the tested cell voltage range up to 2\,V (Figure~\ref{fig:ThvsPE}a).
However, the ThALD-made catalyst showed a higher performance than the PEALD catalyst due to the lower activation overpotential, as indicated by the steep increase of $V_\text{cell}$ at low currents.
Increasing the number of PEALD cycles beyond 150 did not further increase the anode performance (Supporting Information Figure~S6a), likely since the Ir-based layer is already closed at 150 cycles.

To compare the activity of the catalysts in terms of a material property, we performed additional measurements at low currents and fitted the Tafel equation (Equation~\ref{eq:Tafel} in the Methods section) in the low current range up to $10^{-2}$\Acm. 
Even though this yields both the exchange current density $j_0$ and the Tafel slope $b$, we focus the following discussion on the Tafel slope, since $j_0$ is obtained by extrapolating the current density over several orders of magnitude, leading to high uncertainties.\cite{Matsumoto1986}
For the PEALD anode, the observed slope of 66\mVdec is similar to the 72\mVdec observed for crystalline \ce{IrO2} by Reier et al.\cite{Reier2014}, and therefore in line with the observation of \ce{IrO2} via XPS from our PEALD process at 220\,°C.
The Tafel slope of the ThALD-made anode is 51\mVdec, which in the range of reported values of 40--60\mVdec for amorphous IrO$_x$.\cite{Reier2014,Matsumoto1986,Yuan2024}
This agrees with the observation in the XPS discussion (above) that the ThALD-deposited metallic Ir covered by a thin layer of amorphous IrO$_x$.
In addition, metallic Ir typically oxidizes to amorphous IrO$_x$ under anodic potentials.\cite{Siracusano2017}

Note that we limited the fitting range in Figure~\ref{fig:ThvsPE}b to $\leq10^{-2}$\Acm instead of the more commonly used $\leq10^{-1}$\Acm for effective Tafel slopes on the cell level in the literature. 
The logarithmic plots (Figure~\ref{fig:ThvsPE}b) show deviations from the ideal linear behavior already for current densities between $10^{-2}$\Acm and $10^{-1}$\Acm, indicating limitations beyond bare catalyst activity such as mass transport limitations (to be discussed at a later stage).
Limiting the current range to lower currents was a deliberate choice to identify the bare catalyst activity rather than an effective electrode activity.

The Ir loading of both anodes differ dramatically (Table~\ref{tab:anodes}), with 68$\times$ lower loading for PEALD versus ThALD.
During the PEALD process, Ir precursor is dosed while continuously vacuum-pumping the reactor chamber due to technical reasons.
In contrast to this, for ThALD process, the precursor is dosed into the vacuum chamber while pumping is interrupted through a valve between vacuum pump and reactor chamber\cite{Schlicht2018,Schuler2019}. 
This allows for a soaking step to increase precursor utilization, during which the precursor can also diffuse deeper into the Ti felt substrate. 
Thus, the coating reaches deeper into the PTL, where it increases the anode loading, but has no effect on the electrocatalytic activity. 
Furthermore, we suspect that the ozone counter-reactant of the ThALD process may have a longer lifetime than oxygen radicals from the plasma when diffusing into the PTL.\cite{vandePol2025}
This aspect is discussed in more detail in the context of MPL anodes.

\subsection{PEMWE anodes on MPLs}
\begin{figure}
    \centering
    \includegraphics[width=1\linewidth]{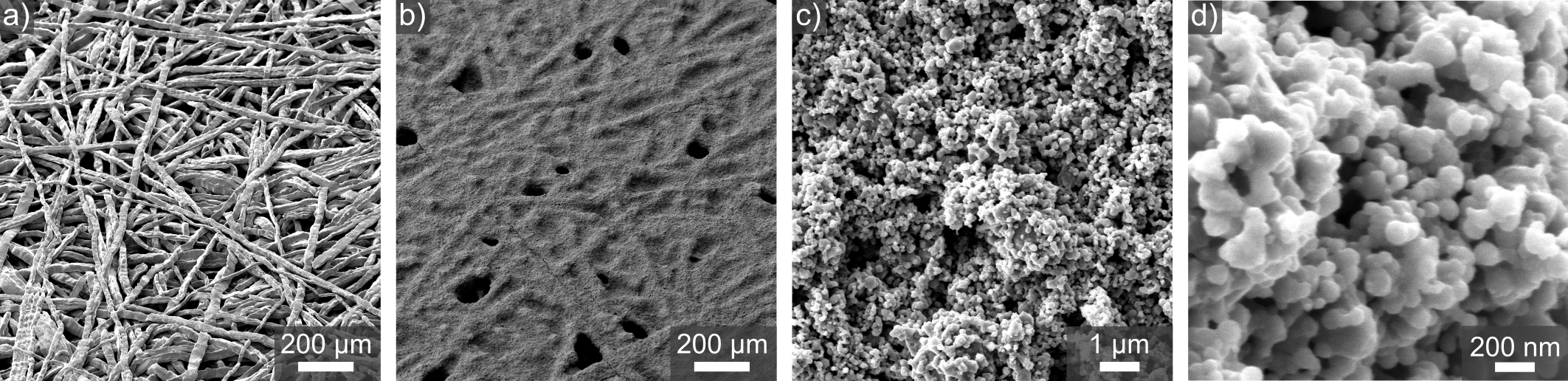}
    \caption{SEM images of the anode substrates used in this paper. a) Bare Ti fiber felt (BF). b)-d) Ti fiber felt coated with Ti MPL from Ti nanoparticles}
    \label{fig:SEM}
\end{figure}

Anodes based on bare Ti felt PTLs offer limited active surface areas which can be increased by adding an MPL.
Bare PTLs consist of a nonwoven mesh of microfibers, and therefore exhibit a rather coarse structure on the microscale (Figure~\ref{fig:SEM}a).
Upon assembly of the membrane with the PTE, only part of the Nafion\texttrademark~membrane surface is in contact with the catalyst coated fiber surface.
Additionally, the fibers exhibit a smooth surface, with a surface area close to their geometrical surface area.
Both aspects contribute to a low active surface area of the PTE for the samples discussed above.
To smoothen the PTL surface on a microscale (Figure~\ref{fig:SEM}b) and at the same time add texture to the fiber surface on a nanoscale (Figure~\ref{fig:SEM}c,d), we developed a spray coating method to equip the PTLs with an MPL (see Methods section for experimental details).
These MPLs were then coated with the same ALD processes as before the bare Ti felts.

\begin{figure}
    \centering
    \includegraphics[width=\linewidth]{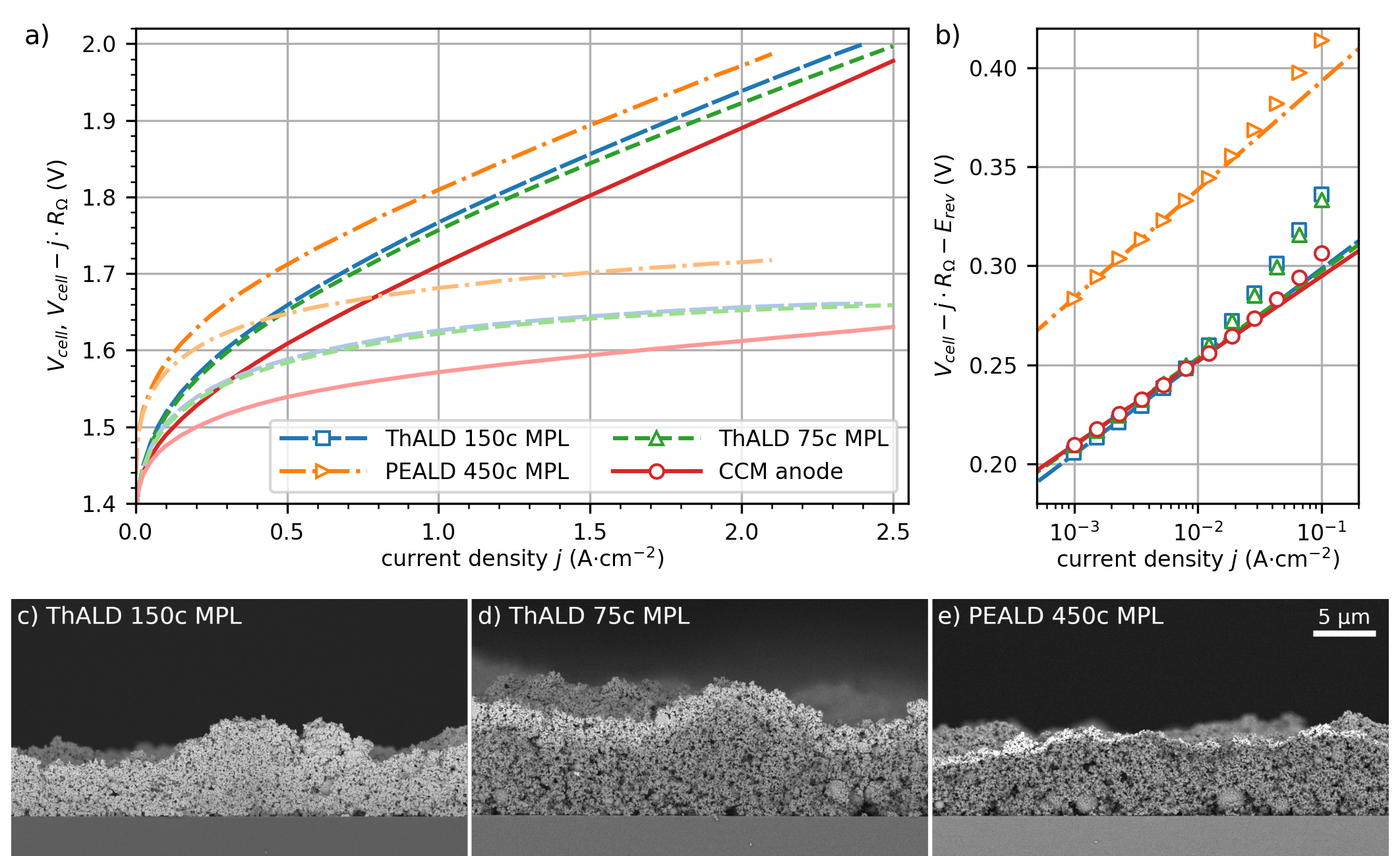}
    \caption{a) Polarization curves for three cells with ALD-made anodes using MPLs and one commercial CCM anode reference. Series resistance-free ($V_\text{cell}-j\cdot R_\Omega$) polarization curves are shown in lighter colors. b) Tafel plots of the same anodes from a separate, low current measurement. Data points are marked with symbols; lines show linear fits in the range from $10^{-3}$ to $10^{-2}$\Acm; fit parameters are given in Table~\ref{tab:anodes}. c--e) Si reference samples, with layer analog to MPL and deposited Ir. Samples were cleaved before recording electron micrographs using a backscattered electron detector (BED) for enhanced material contrast. Ir appears lighter in color, as independently confirmed through EDX linescans (Supporting Information Figure~S10, in this file below). Scaling is identical for figures c--e.
    }
    \label{fig:polwithMPL}
\end{figure}

Anode performances from both the ThALD and PEALD process benefited from the increased surface area of the anode MPL (compare Figure~\ref{fig:ThvsPE}a and \ref{fig:polwithMPL}a).
For the ThALD 150c MPL anode the current increased by about 20\,\% at a benchmark voltage $V_\text{cell}=1.8$\,V compared to the ThALD 150c anode (Table~\ref{tab:anodes}).
In contrast, the PEALD 450c MPL anode more than doubled in current when compared to the PEALD 150c anode on bare felt.

The optimal number of cycles of the PEALD process to achieve the highest performance differs per type of substrate (see optimization in Supporting Information Figure~S6), presumably due to the difference in specific surface area. 
To accommodate for this aspect, we increased the number of cycles from 150 (BF) to 450 (with MPL), while noting that adapting the dosing time may have led to a similar result.

To characterize the catalyst activity, we again performed low-current measurements and obtained the Tafel plots in Figure~\ref{fig:polwithMPL}b.
For consistency, the fitting range was limited to the strictly linear range up to $10^{-2}$\Acm.
The implications of the fitting range will be discussed at a later stage.
For the ThALD process, the Tafel slope decreased moderately by 4\mVdec to 47\mVdec due to the introduction of the MPL (Table~\ref{tab:anodes}), which is similar to literature values for PEMWE anodes\cite{Buhre2023}.
The PEALD MPL anode showed a clearer decrease in Tafel slope by 11\mVdec to 55\mVdec compared to the BF substrate.
This increase in activity in both cases could be related to the nanoscale curvature of the MPL and thus different structure of the Ir and \ce{IrO2} film, respectively. 
It seems plausible that in both cases the curvature led to more lowly-coordinated edge atoms compared to films grown on the smooth microfibers of the BF substrate.\cite{vandePol2025}
However, as the surface of the ThALD Ir was already terminated by an amorphous IrO$_x$ with a non-crystalline structure, the effect in case of the PEALD \ce{IrO2} was stronger, leading to the more pronounced reduction of the Tafel slope.

\begin{table}
    \centering
    \begin{tabular}{cccccc}
         Anode &  \makecell{Ir loading\\\ [mg$\cdot$cm$^{-2}$]} & \makecell{$R_\Omega$\\\ [m$\Omega\cdot$cm$^{-2}$] } & \makecell{Tafel slope $b$ \\\ [\mVdec] } &  \makecell{$j$\,@\,$V_\text{cell}$= 1.8\,V\\\ [\Acm]} \\
         \hline
         ThALD 150c BF&  0.47& 130 & 51 &  0.994\\
         PEALD 150c BF&  0.007& 148 &66 &  0.438\\
         ThALD 150c MPL&  0.82& 141& 47 & 1.180\\
         PEALD 450c MPL&  0.04&128&  55 &  0.949\\
         ThALD 75c MPL&  0.21& 135& 44 &  1.240\\
         CCM anode &  1.4&  139& 43 &   1.491\\
    \end{tabular}
    \caption{Numerical parameters of the cells with anodes shown in Figure~\ref{fig:ThvsPE} and \ref{fig:polwithMPL}. Tafel slopes were obtained by fitting the Tafel equation \ref{eq:Tafel} to the linear range $\leq10^{-2}$\Acm as shown in the respective figure. The series resistance $R_\Omega$ was obtained through fitting an equivalent circuit model (see Figure~\ref{fig:geis}) to EIS data obtained at 0.125\Acm. The Ir loadings were determined via instrumental neutron activation analysis (INAA) on reference samples (Supporting Information Figure~S7).}
    \label{tab:anodes}
\end{table}

The ThALD process was optimized in previous studies\cite{Schlicht2018,Schlicht2018a,Haschke2019}, and further optimization of the process was out of scope for our current study.
However, to still gain insights into the relationship between number of ALD cycles, Ir loading and anode performance, we repeated the ThALD process with 75 cycles.\cite{Schlicht2019}
Remarkably, the loading decreased by 74\,\% (Table~\ref{tab:anodes}, ThALD 150c MPL vs.\ 75c MPL), while the overall performance slightly increased rather than decreased (-3\mVdec Tafel slope, +0.06\Acm current density at 1.8\,V).
While the performance difference is apparent in the polarization curves, the $R_\Omega$-free cell voltage indicates almost identical electrode performances (Figure~\ref{fig:polwithMPL}a). 
This indicates that the cell performance difference was mainly due to the slightly different $R_\Omega$ values (Table~\ref{tab:anodes}).

On a side note, we also tested the ThALD 150c Ir ALD process on Ti frit substrates (Mott Corporation, USA), with and without an added MPL (Supporting Information Figure~S8 and S9).
The performance was overall similar to the Ti fiber felts discussed here, with a slightly more pronounced performance increase due to the MPL.

The relation of anode performance and Ir loading is affected by the Ir distribution within the MPL.
To understand the near-identical performance despite four times lower loading of the 75c anode, we analyzed the cross-sections of Ir-coated MPLs on reference samples via SEM (Figure~\ref{fig:polwithMPL}c--e).
With 150 cycles (c), Ir was deposited throughout the entire MPL thickness of 4 to 8\,µm, whereas for 75 cycles (d), Ir reached only about 2.5\,µm into the MPL. 
These findings area confirmed by EDX cross-section scans (Supporting Information Figure~S10).

Multiple non-exclusive effects could explain the deeper Ir penetration at higher ALD cycle number.
On planar reference samples, we observed a substrate-inhibited nucleation delay (Supporting Information Figure~S4), which points to a significant energy barrier for nucleation of the IrO$_x$ layer. 
Since this barrier is more likely to be overcome at higher local partial pressure of Ir precursor, nucleation occurs at the exposed top surface of the MPL first, and then advances into the MPL at higher number of cycles.
Another possible explanation is the limited lifetime of the ozone counter-reactant.
\ce{TiO2} is known to be a catalyst for ozone decomposition\cite{Ohtani1992}, such that ozone has a high recombination probability when diffusing through the (native oxide) \ce{TiO2} pores.
Thus far, to the best of our knowledge, no literature data on the recombination probability of ozone on IrO$_x$ is available.
However, a lower recombination probability compared to \ce{TiO2} would explain the increasing penetration depth of the ozone counter reactant and thus the of Ir deposition with increasing number of ALD cycles.

Switching from ThALD to PEALD reduces the anode loading by 5$\times$ to 0.04\mgcm, even after 450 cycles.
The MPL reference cross-section image (Figure~\ref{fig:polwithMPL}e) reveals that Ir grows only on the surface of the MPL with $<$\,1\,µm penetration depth.
The lower film penetration depths of radicals from oxygen plasma compared to ozone is a well-known phenomenon from other ALD processes.\cite{Chipmetrics2023,Arts2021,Arts2019}. 
Specifically for a PEALD IrO$_x$ process, Van den Poll et al.\ investigated the penetration depth into a PillarHall substrate, reaching several µm\cite{vandePol2025}. However, this higher depth is likely due to much larger critical dimensions of the test substrate compared to the MPL pores used in the present work.

The discussion of the penetration depth bears the question to which depth the Ir contributes to the catalytic activity.
Given that the ThALD 75c MPL anode gives similar or better activity as the ThALD 150c MPL anode, it is clear that the depth to which Ir contributes to catalytic activity is $\leq2.5$\,µm. 
Comparing the ThALD 75c MPL and the PEALD 450c MPL anode, the latter shows 30\,\% lower currents at 1.8\,V, which could be either: a result of the lower penetration depth or of the lower intrinsic activity of the \ce{IrO2} (Figure~\ref{fig:ThvsPE}). 
Previous results, pointing to the outstanding role of the triple phase boundary\cite{Padgett2023,Yuan2024}, and the fact that the difference in intrinsic activity of PEALD vs.\ ThALD catalyst is quite drastic, suggest that the latter is the main cause for the lower performance of the PEALD anode, and only the catalyst in direct contact with the membrane contributes to the catalytic activity.
This puts low loading (0.04\mgcm) anodes with the same performance as the ThALD 75c MPL anodes within reach, if metallic Ir or higher-activity IrO$_x$ instead of \ce{IrO2} is deposited.\cite{vandePol2025}

We also tested a commercial, single-sided anode CCM with 1.4\mgcm as a benchmark, using the same ionomer-free ALD cathode PTE as for the ionomer-free ALD anodes.
The Tafel slope of the CCM anode is 1\mVdec lower than the best performing ALD anode (ThALD 75c MPL), while there is a notable 
performance difference between the two.
However, despite the 7$\times$ higher Ir loading, the current density at  1.8\,V is only about 20\% higher than for the best performing ALD anode (ThALD 75c MPL, see Table~\ref{tab:anodes}).
Moreover, the current difference becomes even smaller at 2.0\,V (Figure~\ref{fig:polwithMPL}a).
The difference between the two will be analyzed in more detail in the Discussion section.

\subsection{Anode stability}

\begin{figure}
    \centering
    \includegraphics[width=0.75\linewidth]{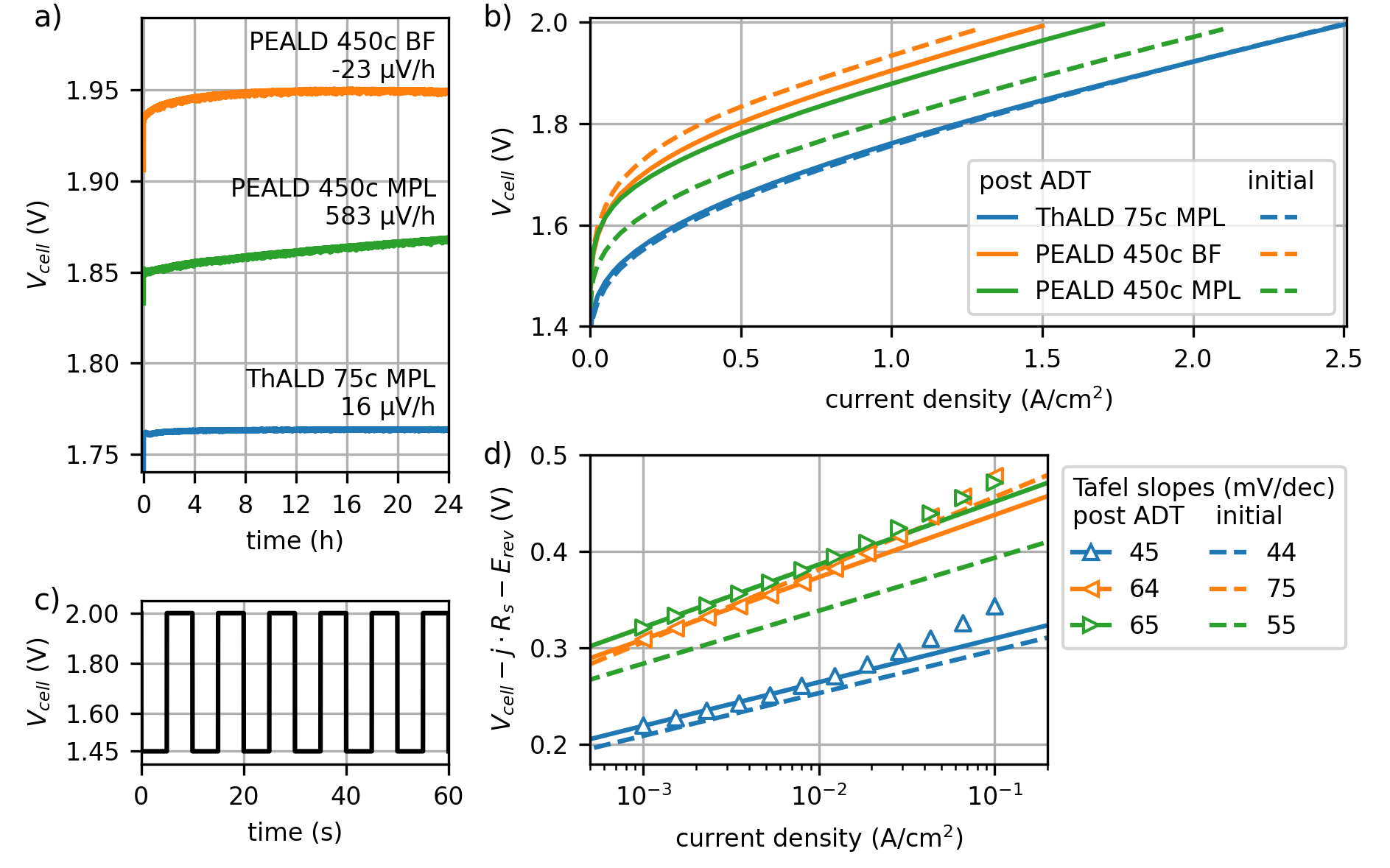}
    \caption{Durability of three representative anodes. a) Evolution of cell voltage $V_\text{cell}$ over 24h under 1\Acm. The presented degradation rates in terms of µV/h cell voltage increase are the result of a linear fit between 12h to 24h. b) Polarization curves before (dashed lines) and after accelerated durability testing (ADT) under dynamic operation. c) Illustration of the driving voltage during ADT. A sequence of 1.45\,V for 5\,s and 2.00\,V for 5\,s was repeated for 25,000 cycles. d) Tafel plot and slopes before (initial, dashed lines) and after ADT (post, solid lines).}
    \label{fig:stability}
\end{figure}

For a first estimate of the stability of our ALD-made anodes, we performed chronopotentiometry measurements at 1\,\Acm over the course of 24\,h and determined the $V_\text{cell}$ increase after an initial break-in period of 12\,h (Figure \ref{fig:stability}a).
Of the three presented anodes, ThALD 75c MPL showed the smallest overpotential increase of 16\uVh. 
The PEALD 450c MPL anode was substantially less stable with $V_\text{cell}$ increasing by 583\uVh.
To learn more about this degradation, we performed the same measurement with an Ir-coated bare Ti felt PTL (PEALD 450c BF), and surprisingly, we find a negative slope (-23\uVh) after an initial increase in $V_\text{cell}$.

To estimate the longer-term stability of our anodes, we subsequently applied an accelerated durability test (ADT), for which alternated $V_\text{cell}$ between 1.45 and 2.00\,V, holding each for 5\,s, for 25,000 repetitions\cite{Alia2019} (Figure \ref{fig:stability}c); this is equivalent to at least 600\,h of intermittent electrolyzer operation, e.g., by wind or solar power. 
After the ADT procedure, we again recorded the polarization curves (Figure \ref{fig:stability}b) and, separately, the low-current Tafel slope characterization (Figure \ref{fig:stability}d).
The ThALD 75 MPL polarization curve shows only minor deviations from the initial (beginning of life) polarization curve (dashed line underneath solid line).
The main difference is the slightly higher $V_\text{cell}$ at low currents due to the 1\mVdec increase in the Tafel slope (Figure \ref{fig:stability}d).
For PEALD 450c MPL, the ADT measurement confirms the trend of the 24\,h measurement, with severe performance decrease due to activity loss indicated by the 10\mVdec increase in Tafel slope. 
In contrast, the PEALD 450c BF anode again showed a significant performance increase indicated by a decrease in Tafel slope by -11\mVdec.
Moreover, even during the first galvanostatic polarization curve recordings of PEALD anodes fabricated on bare Ti felts (without MPL), we consistently observed a jump in performance at around 1.8\,V (Supporting Information Figure~S11).

A potential reason for the performance increase of the PEALD 450c anode is a qualitative change to the \ce{IrO2} catalyst during operation. 
To investigate this possibility, analyzed the anode using high resolution XPS and compared it to a \ce{IrO2}-coated Ti felt that was not tested in the PEMWE cell.
Indeed, we find a partial transformation of \ce{IrO2} to IrO$_x$ with a different oxidation state, or to IrOH (Supporting Information Figure~S12)\cite{Freakley2017}.
It is therefore plausible that part of the performance increase is due to this transformation to higher-activity IrO$_x$\cite{Reier2014,Matsumoto1986,Yuan2024}.
Similarly, Siracusano et al.\ observed a partial transformation of stoichiometric \ce{IrO2} into sub-stoichiometric IrO$_x$ in used anodes via XPS; however, their tests over 1000\,h showed a cell voltage increase rather than a decrease which could be due to a loss of catalyst.\cite{Siracusano2017}
In addition to the change of the catalyst, we observed in previous experiments that the fibrous surface of the PTL indents into the membrane, leading to more contact area between catalyst and membrane.
Here, we cannot exclude the possibility that this effect increases over time, thus gradually lowering cell voltages and Tafel slopes.

The cause of the performance loss of the PEALD 450c MPL anode is more difficult to investigate.
The MPL and with it the catalyst layer stuck to the Nafion\texttrademark~117 membrane upon disassembly of the cell, therefore the used catalyst was not accessible to post-testing XPS analysis.
We can only stipulate that the degradation is due to the interplay of the PEALD process with the MPL, as neither the ThALD process on MPL nor the PEALD process on bare Ti felts (BF) shows a similar degradation.
As the plasma process led to a notable oxide growth in the underlying Ti (Supporting Information Figures~S3) in planar substrates, we assume that this was also the case in the nanoparticle-based MPL. 
If and how this can lead to increased catalyst degradation should be addressed in future research.

\section{Discussion}

In the following, we discuss three main topics: First, we study in depth the qualitative differences between an ionomer-free ALD anode on MPL and the CCM references through a reconstruction of overpotential contributions from EIS data. Second, we discuss the powder-specific Ir loading, current density and voltage efficiency for a range of anodes. Third, as an outlook, we demonstrate the versatility of ionomer-free anodes by combining them with AEM-based \ce{CO2} electrolysis.

\subsection{Reconstruction of overpotential contributions via EIS}

The near-identical Tafel slopes of the CCM anode and ThALD 75c MPL anode suggest similar activation losses, yet the polarization curve of ThALD 75 MPL shows up to 55\,mV higher $V_\text{cell}$ (Figure~\ref{fig:polwithMPL}).
This difference is similar in the $R_{\Omega}$-free polarization curves, indicating higher activation losses in the ThALD 75c MPL anode than the Tafel analysis suggests.
Arbitrarily increasing the fitting range beyond the linear range in Figure~\ref{fig:polwithMPL}b would increase the Tafel slope, but would obscure rather than solve the contradiction.\cite{parkCorrelationTafelAnalysis2019}
However, identifying the activation overpotential is an important step in the full overpotential breakdown;
as we will show, this is particularly useful to identify qualitative differences between a ionomer-free, ALD-made anodes and CCMs.\cite{BerntIonomer2016}
In the following, we therefore discuss an alternative method of overpotential breakdown which is based on EIS, and therefore devoid of arbitrarily selecting the Tafel fit range.\cite{parkCorrelationTafelAnalysis2019}

\begin{figure}
    \centering
    \includegraphics[width=1.0\linewidth]{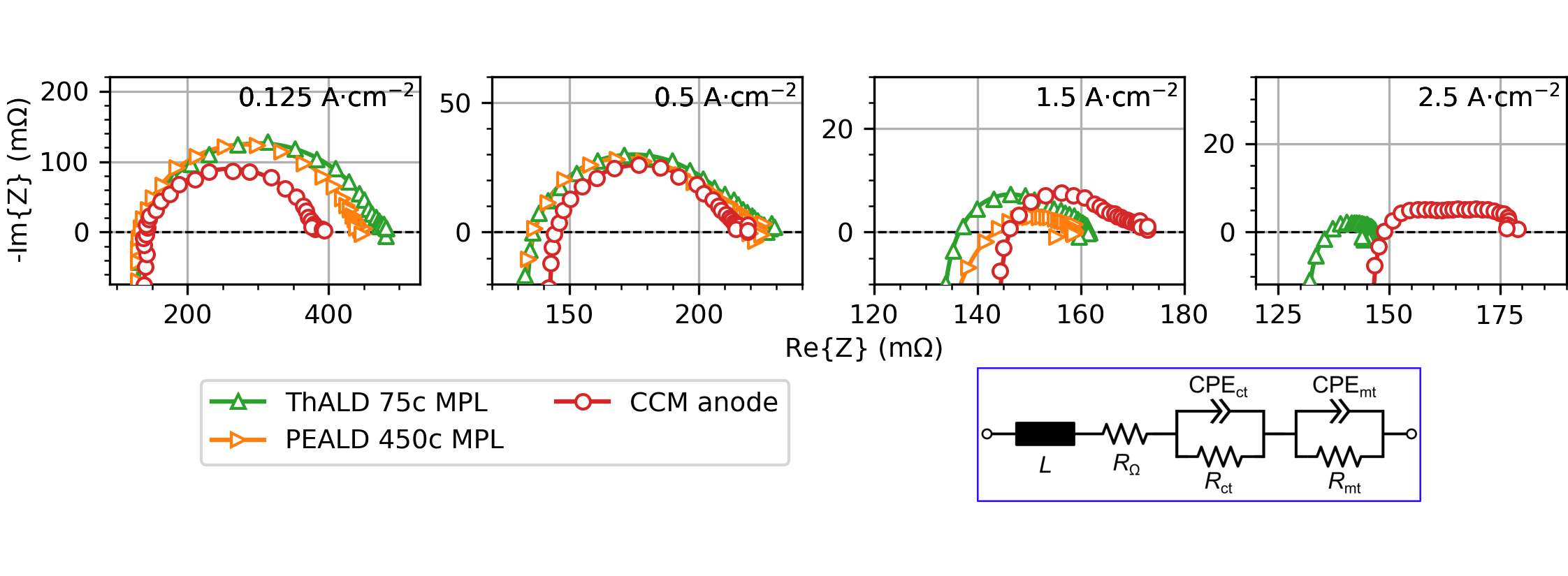}
    \caption{Nyquist plots from EIS data for three representative anodes, and four representative current densities (note that PEALD 450c MPL reached only 2.1\Acm). The equivalent circuit model (blue box) was fitted to the experimental data, and included the inductance $L$, series resistance $R_{\Omega}$, and Voigt elements\cite{GarciaNavarro2019,Dedigama2014} for charge transfer $ct$ and mass transport $mt$.}
    \label{fig:geis}
\end{figure}

We performed electrochemical impedance spectroscopy measurements for three representative anodes -- ThALD 75c MPL, PEALD 450c MPL and CCM anode -- at DC operating points in the range 0.125--2.5\Acm (Figure~\ref{fig:geis}).
The fitted equivalent circuit model includes the inductance of the wiring $L$, the series resistance $R_{\Omega}$, a Voigt element for charge transfer (CPE$_\text{ct}||R_\text{ct}$) and another Voigt element for mass transport (CPE$_\text{mt}||R_\text{mt}$)\cite{GarciaNavarro2019,Dedigama2014}.
\cite{Kim2022}
Note that both the anode and the cathode charge transfer resistances are included in a single Voigt element.
Since the anode is expected to be the dominant part\cite{BerntIonomer2016}, and moreover, the cathodes are identical for all discussed anodes, this simplification will not affect the outcome of the discussion.
For a cathode-focused work, EIS measurements with more densely-spaced frequencies (Figure~\ref{fig:geis}) especially at high frequencies, could help to dissect the anode and cathode contributions to charge transfer resistance\cite{Siracusano2018}.

To gain an overview, we start with qualitative observations on the Nyquist plots in Figure~\ref{fig:geis}.
At 0.125\Acm, the strongest contribution for all three anodes is the charge transfer resistance, respectively, indicated by a single arc.\cite{Lazanas2023}
The curve of the CCM anode intersects the Re\{Z\} axis at a lower value, equivalent to a lower charge transfer resistance. 
At 0.5\Acm, all three charge transfer resistance-related arcs show some distortion toward low frequencies (right intersection with the Re\{Z\} axis).
This distortion develops into an additional arc at 1.5\Acm for ThALD 75c MPL and CCM anode, which points to mass transfer resistances\cite{Lazanas2023}, which is less pronounced for PEALD 450c MPL.
At 2.5\Acm this trend continues for the CCM anode but ThALD 75c MPL reverts to a single arc.

\begin{figure}
    \centering
    \includegraphics[width=0.5\linewidth]{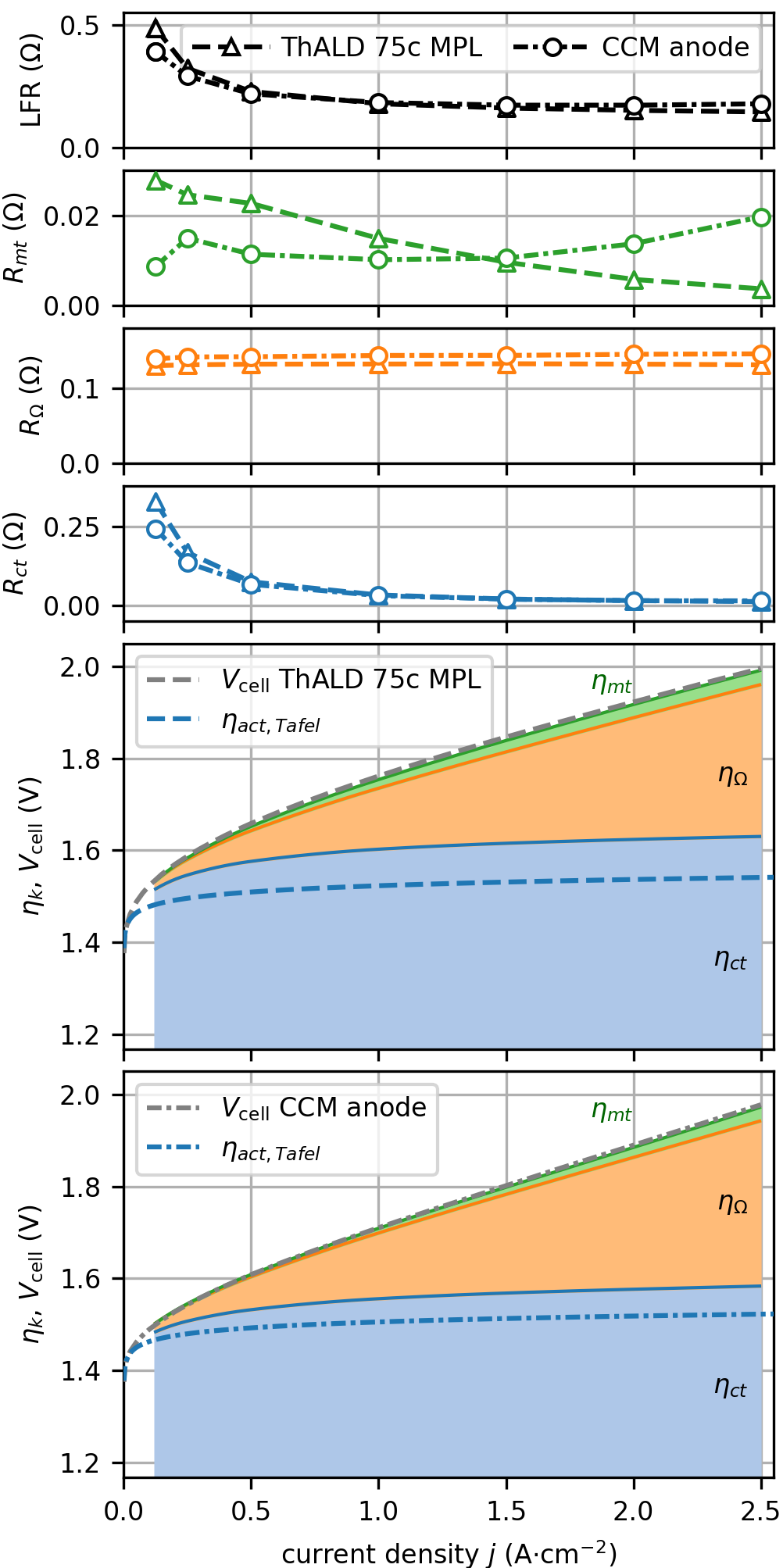}
    \caption{Overpotential reconstruction from EIS data. Top: low frequency resistance (LFR) decomposed into components $R_k$ from mass transport ($R_\text{mt}$), ohmic ($R_{\Omega}$) and charge transfer ($R_\text{ct}$), obtained via fitting equivalent circuits to EIS data, respectively. Bottom: overpotential contributions $\eta_k$ from integrating $R_k$, and comparison to $V_\text{cell}$ from an independent measurement. Markers and linestyles represent samples, colors represent contributions $k$.}
    \label{fig:Etareconstruct}
\end{figure}

Recognizing that EIS is a differential method -- it identifies the impedance contributions to a change in $V_\text{cell}$ at the respective operating point -- we can use EIS data to reconstruct the polarization curve from its individual overpotential contributions.
The rightmost intersection of the impedance curve (Figure~\ref{fig:geis}) with the Re\{Z\} axis is the low frequency resistance (LFR), which is the sum $\sum_k R_k$ of mass transport ($R_\text{mt}$), ohmic ($R_{\Omega}$) and charge transfer ($R_\text{ct}$) contributions (Figure~\ref{fig:Etareconstruct}).
Moreover, in the limit of low frequencies, the slope of the polarization curve is identical to the impedance, and can be estimated from the LFR\cite{Wallnofer2024}:
\begin{equation}
Z(\omega\rightarrow0)=\frac{\text{d}V_\text{cell}(j)}{\text{d}j}\approx\text{LFR}(j).    
\end{equation}
Therefore, integration yields the polarization curve\cite{Lihaokun2023},
\begin{equation}\label{eq:integralLFR}
V_\text{cell}(j)\approx\int_0^j \text{LFR}(j)\text{d}j+\text{const.}=\sum_k\int_0^j R_k(j)\text{d}j+c_k
\end{equation}
where the sum is the breakdown of the polarization curve into its individual overpotential contributions.
The integration constants $c_k$ can be calculated from a potentiometric measurement at 0.125\Acm obtained during EIS measurements, and using two assumptions: 1) mass transport resistances are small compared to the other contributions at small currents (Figure~\ref{fig:Etareconstruct}), and will be even smaller for $j\rightarrow0$.
The integral $\int R_\text{mt}\text{d}j$ is therefore negligible at currents $\leq$0.125\Acm, and therefore $c_\text{mt}=0$.
2) The ohmic resistance is constant, and therefore $c_{\Omega}=j\cdot R_{\Omega}$. 
Then the last integration constant for charge transfer is $c_\text{ct}=V_\text{cell}(j$=0.125\Acm$)-E_\text{rev}-c_{\Omega}$, where $E_\text{rev}$ is the reversible potential for water splitting at 80\,°C.

As basis for the polarization curve reconstruction, we complemented the EIS data (Figure~\ref{fig:geis}) with three additional DC operating points (Supporting Information Figure~S13), and linearly interpolated the $R_k$ (broken lines in Figure~\ref{fig:Etareconstruct}, top).
The integration in Equation~\ref{eq:integralLFR} yields the reconstructed polarization curves for ThALD 75c MPL and the CCM reference, and the result is shown in Figure~\ref{fig:Etareconstruct} (bottom) along with polarization curves obtained through independent, direct measurements (broken gray lines). 
Note that the reconstruction of the overpotentials is entirely independent of the polarization curve measurement. Yet, the deviation between the reconstruction and the direct measurement is $<$10\,mV throughout the current range, which we deem sufficient for the sake of the discussion of the relative overpotential contributions $\eta_k$. 
A key requirement for this accurate reconstruction is electrode stability; in the case of the PEALD MPL electrodes, this is not given, and these electrodes are therefore excluded from this analysis.

\subsubsection{Ohmic resistance}
The ohmic resistance $R_\Omega$ is approximately constant for both anodes, and higher for the commercial CCM anode. 
The CCM anode catalyst layer consists of a mixture of ionomer binder and catalyst particles, and is known to add to both the protonic resistance and electronic resistance\cite{Grigoriev2017,BerntIonomer2016}.
The MPL of the ThALD anode, in contrast, consists of sintered metal, and likely exhibits a lower electronic resistance. 
Also, no more than the Ir-coated part facing the membrane contributes to the reaction, thereby the protonic resistance contribution to $R_\Omega$ is lower than for the CCM.
We note, however, that the commercial CCM has undergone processing steps beyond our control, potentially changing the Nafion\texttrademark~117 (N117) membrane resistance compared to the bare membrane used for ALD-made PTEs.

On a side note, we found that the apparent high frequency resistance (HFR, intersection of impedance curve with Re\{Z\} axis) can differ notably from $R_\Omega$ obtained through fitting by up to 10\mOcm. 
The higher apparent HFR value in the latter is explained by the inductance of the cables which effectively increase the resistance at high frequency.
The fitted model, in contrast, accounts for the inductance and therefore provides a more accurate series resistance $R_\Omega$ rather than the HFR.

\subsubsection{Charge transfer}
The charge transfer resistance $R_\text{ct}$ is slightly higher for the ThALD 75c MPL anode, especially for low current densities. 
Due to the integration (Equation~\ref{eq:integralLFR}), this slight difference leads to notably higher overpotential of the ThALD 75c MPL anode, and is therefore at the core of the lower performance (compare $\eta_\text{ct}$ at 2.5\Acm around 1.6 V for both anodes).

A widely employed way to characterize the activation overpotential $\eta_\text{act,Tafel}$ (Equation~\ref{eq:Tafel} in Methods section) is by extrapolating the Tafel-like behavior from low currents -- typically 10$^{-3}$\Acm$\leq j\leq$$^{-1}$\Acm -- to the full current range of the polarization curve, which is additionally plotted in Figure~\ref{fig:Etareconstruct}\cite{Weber2023,Schuler2019,Padgett2023}.
Though activation overpotentials from charge transfer resistance $\eta_\text{ct}$ or (alternatively) from Tafel analysis $\eta_\text{act,Tafel}$ are rooted in the same underlying physics,\cite{parkCorrelationTafelAnalysis2019} the latter yields lower overpotentials by several 10\,mV for both anodes, with higher differences for the ThALD 75c anode. 
To quantify the differences, we determined an effective Tafel slope by fitting the Tafel equation to $\eta_\text{ct}$ (see Supporting Information Figure~S14 for fitted curves), and obtain $b_\text{eff}$=84\,mV for the ThALD 75c MPL anode, -- instead of $b=45$\,mV from low-current logarithmic fit -- and $b_\text{eff}$=76\,mV for the CCM anode -- instead of $b=43$\,mV. 

The concept of distinguishing $b$ and $b_\text{eff}$ in porous electrodes has been described from a theoretical perspective\cite{Kulikovsky2014,Haverkort2019}, but is rarely considered in the PEMWE literature.\cite{Padgett2023}
Haverkort et al.\cite{Haverkort2019}\ found that very dissimilar ion- and electron conductivity within a porous catalyst layer can lead up to doubling the Tafel slope, i.e., $b_\text{eff}/b\approx2$, showing a transition from $b$ to $b_\text{eff}$ at low current densities.
For ThALD 75c MPL, this transition is apparent at around $10^{-2}$\Acm (Figure~\ref{fig:polwithMPL}b), and $b_\text{eff}/b\approx1.87$.
In the Tafel slope of the CCM anode, the transition is only apparent upon close inspection.\cite{Schuler2019,BerntIonomer2016,Bernt2018}

\begin{figure}
    \centering
    \includegraphics[width=0.6\linewidth]{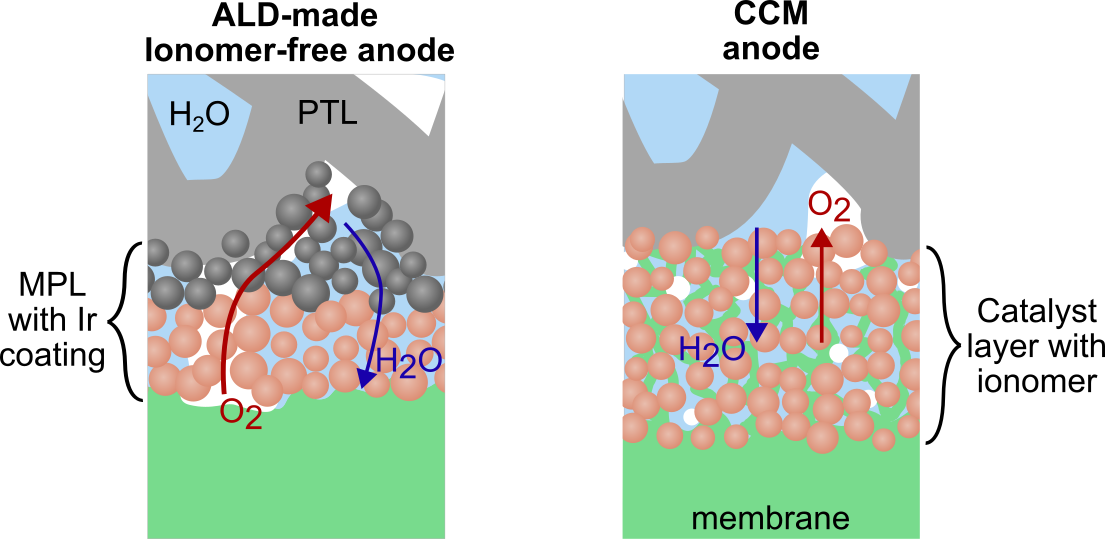}
    \caption{Proposed pathways of water and \ce{O2} transport in MPLs with ALD Ir coating compared to conventional CCM anodes. In ALD-made anodes (left), no ionomer is present within the Ir-coated MPL. The OER occurs primarily at the membrane-MPL interface and gaseous \ce{O2} forms channels through the MPL from the onset of the OER, and therefore lead to mass transport resistance at low currents. The catalyst layer of the CCM anode, in contrast, contains ionomer. Liquid water- and \ce{O2} transport are competing phenomena in opposite direction, where increasing OER leads to poor percolation of liquid water at high currents.}
    \label{fig:mtdiscussion}
\end{figure}

Indeed, the Ir-infiltrated MPL of the ThALD 75c MPL anode may be considered a porous catalyst layer with high electron conductivity due to metallic sinter structure and low ion conductivity only provided by water. 
However, a strong contribution to catalytic activity is expected from the catalyst that is in direct contact with the membrane (schematic illustration in Figure~\ref{fig:mtdiscussion}, left), as clearly observed for the PEALD 450c MPL anode.
In addition, we suspect that the rather localized reaction at the membrane-MPL interface may lead to gaseous oxygen bubbles, potentially blocking some of the catalyst surface, as further discussed in the context of mass transfer (below).
Paralleling the optimization of catalyst layer thickness in Ref.\ \citenum{Haverkort2019}, future theoretical work could cover the optimization of the MPL structure and catalyst distribution in ionomer-free ALD-made electrodes, and also include mass transport phenomena for the specific case, in order to increase catalyst utilization\cite{Schuler2019}.

\subsubsection{Mass transport}
The opposing trends in mass transport resistance (Figure~\ref{fig:Etareconstruct}) suggest a qualitative difference in the mass transport regimes in the ThALD 75c MPL anode and the CCM anode.
In the CCM anode at low currents, oxygen is transported out of the catalyst layer (cross-section SEM in Supporting Information Figure~S15) in gaseous form or by diffusion through the ionomer (Figure~\ref{fig:mtdiscussion}, right). 
Only at high currents, the out-flowing gas increasingly hinders liquid water inflow, leading to local water depletion.
This is notable in the form of increasing mass transport resistance with increasing current (Figure~\ref{fig:Etareconstruct}, top).
For the ThALD 75c anode, or generally ALD-made anodes of the same structure, the triple phase boundary\cite{Padgett2023,Yuan2024} and OER are strongly localized at the catalyst-membrane interface (Figure~\ref{fig:mtdiscussion}, left).
As the current density increases, the gas pressure in pockets (bubbles) near this interface eventually overcomes the capillary pressure of the fully wetted MPL, and channels are formed through which the oxygen can escape. 
Therefore, $R_\text{mt}$ is high in the ThALD 75c anode for low currents, and decreases the higher the current (Figure~\ref{fig:Etareconstruct}, top).
\begin{figure}
    \centering
    \includegraphics[width=0.5\linewidth]{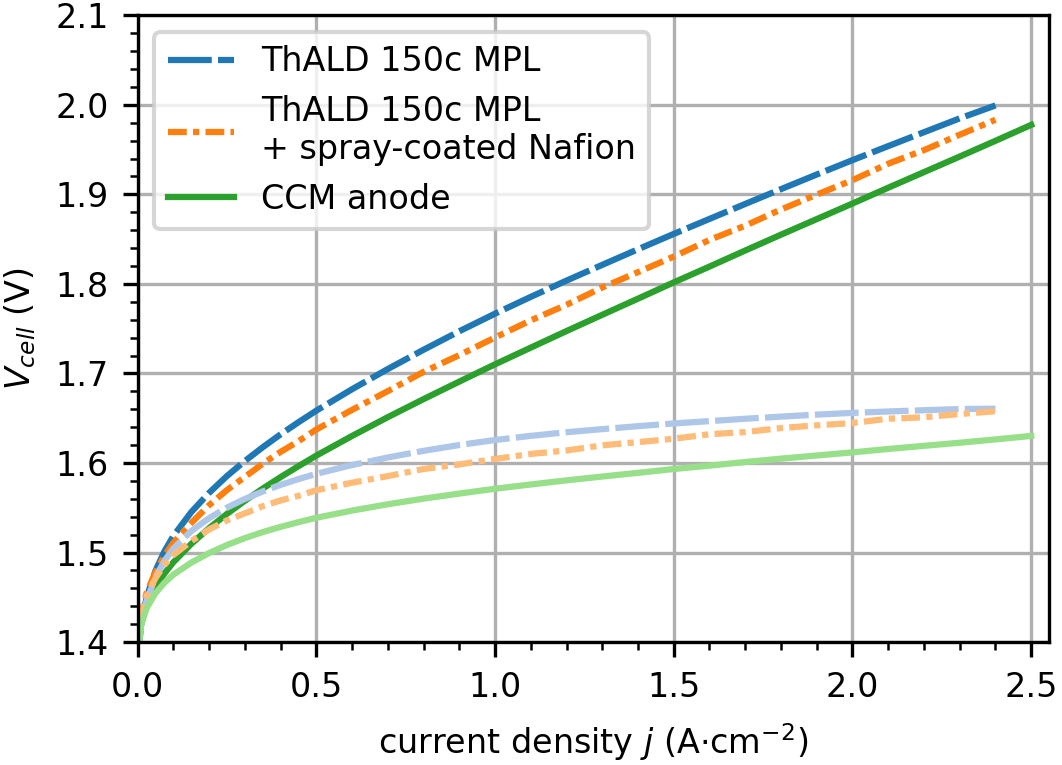}
    \caption{Polarization curves with an additional layer of spray-coated Nafion\texttrademark~on top of the MPL catalyst layer in comparison to the previously shown ThALD 150c MPL anode and commercial CCM anode reference. $R_{\Omega}$-free polarization curves are show in lighter colors.}
    \label{fig:XtraNafion}
\end{figure}

To test this hypothesis, we spray coated ionomer solution onto a 150c ThALD MPL anode to infiltrate the Ir-coated MPL (cross-section SEM in Supporting Information Figure~S16), and thereby shifted the triple phase boundary into the MPL, while also increasing the ionomer-catalyst interface area.
Indeed, the polarization curve is shifted towards CCM polarization curve with an overall increase in current (Figure~\ref{fig:XtraNafion}). 
At the same time, the slope of the $R_{\Omega}$-free polarization curve in the limit of high currents increases, now more closely resembling the slope of the CCM curve, which supports our hypothesis of a transport mechanism as depicted in Figure~\ref{fig:mtdiscussion}.

\subsection{Power-specific loading, current density \& voltage efficiency}

\begin{figure}
    \centering
    \includegraphics[width=0.45\linewidth]{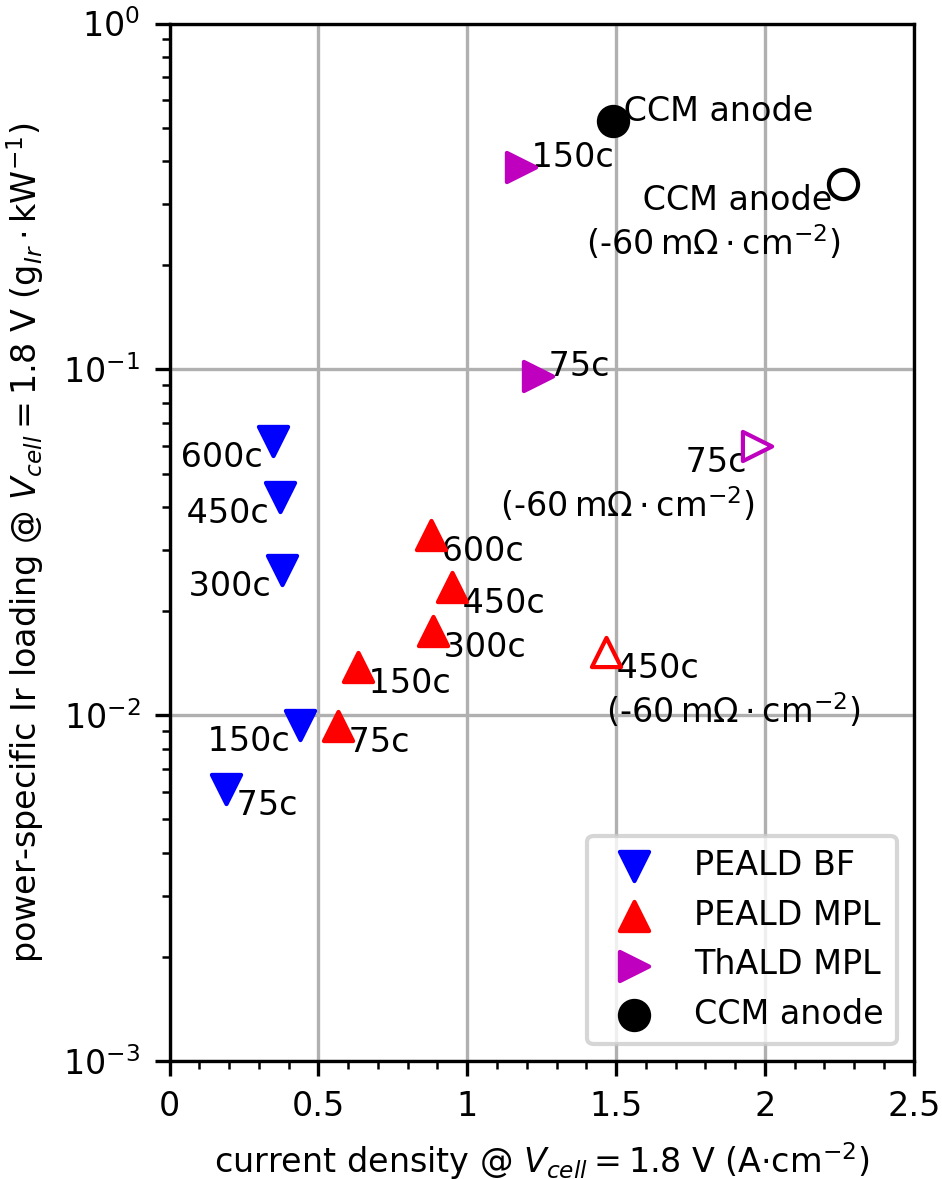}
    \caption{Power-specific mass loading versus current density at $V_\text{cell}=1.8$\,V and $\approx70\,\%_\text{LHV}$ efficiency for anodes tested with Nafion\texttrademark~117 membranes (filled symbols), and calculated values of anodes with a hypothetically thinner membrane (open symbols). The latter were calculated analogous to $R_{\Omega}$-free polarization curves, but subtracting $j\cdot60$\mOcm instead of $j\cdot R_{\Omega}$.}
    \label{fig:massspecific}
\end{figure}

Thus far the specific properties of low-Ir-loading ALD-made anodes with and without MPL have been discussed from the bottom up.
Changing the perspective to a top-down view, we next discuss the low loading of multiple anodes in the context of other performance metrics.\cite{Bernt2018}
First, we consider the conversion efficiency from electrical to chemical energy, which reflects the operational cost of electricity for hydrogen generation.
The efficiency is directly linked to the cell voltage,\cite{Solten2010,Bernt2018} such that $V_\text{cell}=1.8$\,V at 80\,°C is equivalent to an efficiency of $\approx70\,\%_\text{LHV}$ versus lower heating value (LHV) of hydrogen.
Since this is a typical value in technology roadmaps and academic literature\cite{GreenHydrogenCost2020,Bernt2018}, we specify $V_\text{cell}\equiv1.8$\,V to simplify the following discussion.
Second, to reflect the limited Ir supply when rolling out PEMWE at the GW scale, we specify the Ir loading as power-specific, in units \gkW.
Third, we identify the current density for a given anode, which scales approximately inversely proportional to the capital expenditure per electrolyzer capacity.
This thought-process results in the performance overview for multiple anodes in Figure~\ref{fig:massspecific}.

Reducing the Ir loading to meet typical goals\cite{GreenHydrogenCost2020,Wang2025} in the range of 0.1--0.4\gkW appears straightforward, but simultaneously achieving high currents is challenging. 
Considering the PEALD anodes, lowering the number of cycles and thus the loading (Supporting Information Figure~S7) consistently reduces the power-specific Ir loading, with values consistently below the CCM anode.
Increasing the number of cycles increases the current density at first, but reaches its maximum at 150c for bare felt substrates (PEALD BF), and at 450c for the MPL substrates (PEALD MPL), respectively.
A similar optimization of the number of ThALD cycles was out of scope of the present study. 
However, based on a previous optimization of the same process,\cite{Schlicht2019} and the similar current densities for ThALD MPL 75c and 150c, we conclude that the performance in terms of current density reached its maximum for the combination of our ThALD process and our custom-made MPLs. 

Though comparable, the current density of ThALD MPL 75c stays below the CCM anode, while the near-identical Tafel slopes (Figure~\ref{fig:polwithMPL}) indicate similar catalytic activity. This suggests that the way to increase the current density of ThALD MPL 75c is by advancing the design of the MPL, for which literature holds numerous suggestions\cite{Ernst2022,Ernst2024,Kang2018,leePioneeringMicroporousLayers2024}.

The series resistance $R_{\Omega}$ is the dominant contribution to overpotential increase\cite{Suermann2015}, starting from current densities $>$1\Acm (see $\eta_{\Omega}$ in Figure~\ref{fig:Etareconstruct}).
In the present study we tested our anodes with relatively thick N117 membranes which exhibit relatively high resistances.\cite{Sun2019}
To estimate the performance in combination with thinner membranes, we calculated polarization curves with reduced series resistance, $V'_\text{cell}= V_\text{cell}-j\cdot 60$\mOcm, similar to the calculation of $R_{\Omega}$-free polarization curves, and again identified the current density at $V'_\text{cell}=1.8$\,V. Note that this current density was measured experimentally, and only the value for $V'_\text{cell}$ was calculated.
The here chosen $-60$\mOcm correspond to the typical difference between an N117 and a thinner, N212 membrane\cite{Sun2019,Xie2024}.
For a hypothetical configuration of a such thin membrane with the 450c PEALD MPL anode or the 75c ThALD MPL anode, this would increase the current density by 54\% or 59\%, respectively  which also lowers the power-specific Ir loading (Figure~\ref{fig:massspecific}). However, a similar performance increase is expected for the CCM anode.

\subsection{Outlook: AEM-based \ce{CO2} electrolysis}

Traditional wet-deposited catalyst layers include an ionomer binder matrix in which the catalyst particles are embedded.
Thus, they can only be combined with a given membrane type if the ionomers of membrane and catalyst layer are compatible.
In contrast, ionomer-free electrodes can be combined with a range of different membrane materials and technologies.

\begin{figure}[ht]
    \centering
    \includegraphics[width=0.5\linewidth]{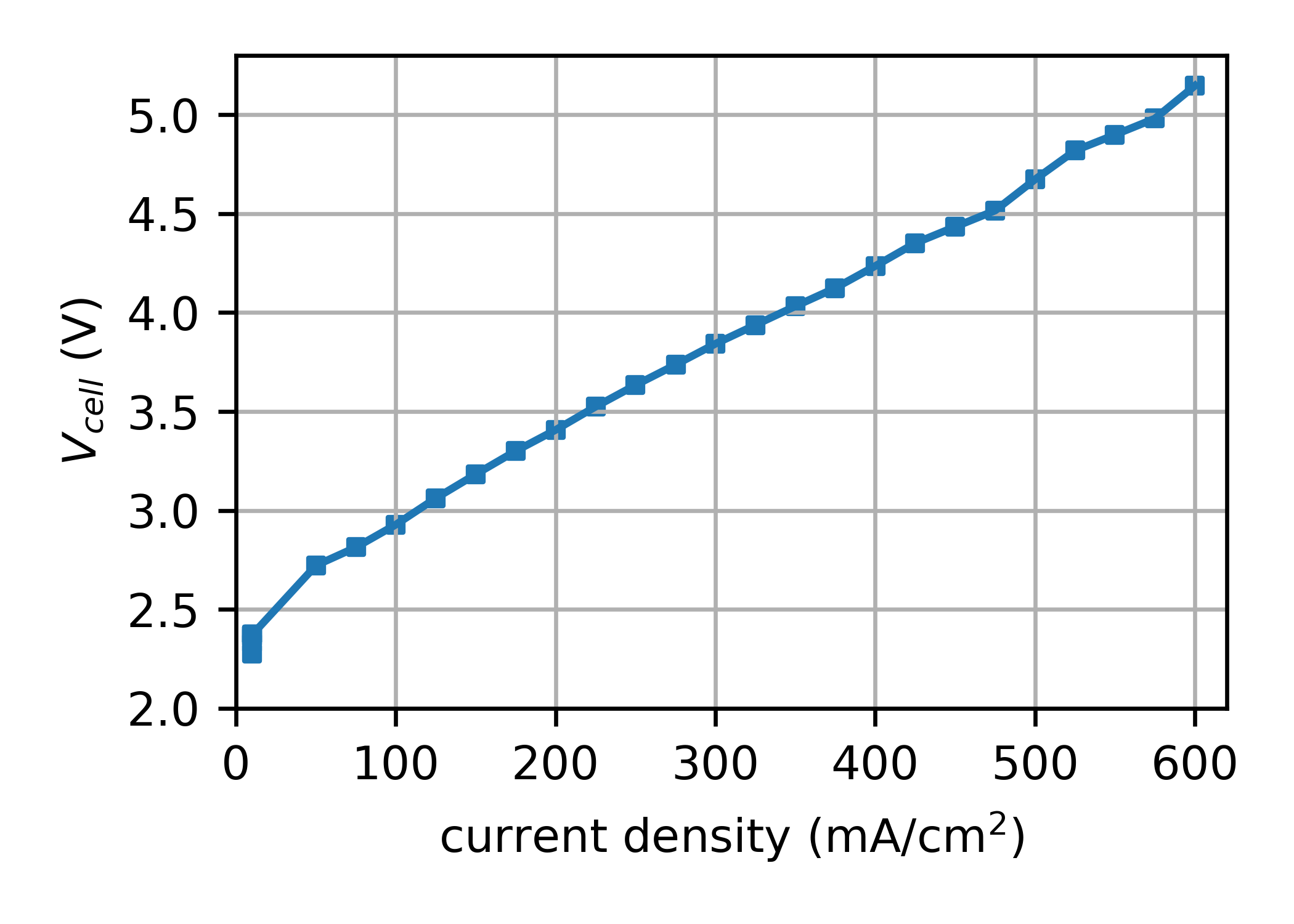}
    \caption{Polarization curve of an AEM-based \ce{CO2}-to-CO electrolysis experiment, using a ThALD 150c MPL Ir anode, and an Ag cathode that was obtained by sputtering (see Methods section.}
    \label{fig:co2rr}
\end{figure}

To demonstrate the versatility of the ionomer-free, direct-ALD catalyst deposition approach, we tested an iridium PTE anode (ThALD 150c MPL) in an AEM-based \ce{CO2} electrolysis experiment across a current range of 10 to 600\mAcm. 
The initial voltage at 10\mAcm was measured at 2.28\,V which closely approaches the reported feasible minimum of 2.16\,V for alkaline \ce{CO2} electrolysis to CO \cite{Schmid2023}, confirming the efficacy of the PTE anode. 
Between 100 and 600\mAcm, the cell voltage increases linearly from 2.93 to 5.15\,V. 
This linearity indicates that ohmic cell resistance, rather than anode catalyst-specific losses, dominates the voltage increase \cite{Hansen2022}. 
These results establish the viability of the ALD-made ionomer-free anode for \ce{CO2} electrolysis systems at higher current densities. 
While optimizing gaskets and cell compression will further decrease ohmic resistance, this experiment clearly demonstrates the immediate applicability of ALD-made ionomer-free electrodes in a \ce{CO2} setup.

\section{Conclusion}

For the first time, ALD-made, ionomer-free anode PTEs were demonstrated to achieve comparable current densities to standard, commercial CCMs, at an Ir loading of 0.21\mgcm and thereby 7$\times$ lower than the CCM reference. 
Moreover, since we used ALD-made ionomer-free cathode PTEs\cite{Chen2026} with even lower Pt loading of 22\ugcm, the total PGM loading was only about 0.24\mgcm.

ThALD led to Ir/IrO$_x$ with higher activity, while PEALD yielded lower activity \ce{IrO2}.
Using an MPL substrate with higher surface area for PTE fabrication reduced this difference, but the MPL did not fully compensate for the lower intrinsic activity of \ce{IrO2}.
Moreover, the interplay of the MPL with the PEALD process led to poor stability of the \ce{IrO2} catalyst. 
However, through PEALD, we effectively tuned the Ir distribution within the MPL and thereby reduced the Ir loading to 0.04\mgcm. 
Further investigations of other ALD reactant sequences could help combine the active and stable IrO$_x$ catalysts from ThALD with the low loading of PEALD, e.g., by using a different counter-reactant than oxygen plasma, or adding a reductive step such as \ce{H2} plasma exposure.\cite{Chen2026}

The catalyst loading is easily controllable through the number of ALD cycles but the activity reaches its maximum once a given substrate surface is fully covered by Ir.
Higher activities can be achieved by increasing the surface area of the substrate, as the example of coating bare Ti felts versus MPLs shows.
Independent of catalyst deposition, PTL substrates could be tailored to combine a high surface area facing the membrane, high electric conductivity and high mass transport. 
Literature provides ample inspiration for innovative PTL designs, ranging from Ti sinter substrates\cite{Ernst2022,Ernst2024,leePioneeringMicroporousLayers2024} and combinations of such with custom substrates\cite{Kang2018} to novel, unconventional substrate types like inverse opals.\cite{Moehle2026,sprafke2013}
ALD enables tailoring the distribution of IrO$_x$ specifically to these PTLs, as demonstrated in this work for the case of spray-coated MPLs.

Beyond PEMWE, we showcased the applicability of ionomer-free electrodes to AEM-based \ce{CO2} electrolysis.
Besides demonstrating the versatility of the tested Ir anodes, it also underlines the potential of ALD to fabrication of ionomer-free electrodes more generally. 
Beyond simple oxides, ALD enables a plethora of multi-component compounds that can be tailored to the desired anodic or cathodic half-reaction.

\subsection{Methods}
\subsubsection{Substrate preparation}
Commercial Ti fiber felts with 1 mm thickness, 20\,µm fiber diameter and 56\,\% porosity were acquired from Bekaert (Belgium).
For use as anode PTLs, $2\times2$\,cm$^2$ samples were laser-cut from larger sheets.
As reference samples for Ir loading determination via INAA and microstructure analysis, smaller $0.5\times1$\,cm$^2$ samples were manually cut using Ti scissors. 
All samples were first rinsed in isopropyl alcohol (IPA), then cleaned in an ultrasonic bath of IPA and dionized water subsequently for 20\,min each, before drying at 80\,°C in air.

For the MPL anodes, an ink was prepared by first mixing Ti nanoparticles (30-80 nm, 99.9\,\% Ti, Nanografi, Turkey) with a solution of methylcellulose (M0512, Sigma Aldrich), IPA and deionized water. 
After ultrasonication for 15 min, ethylene glycol was added, and the ink sonicated again for another 60 min.\cite{Kang2018}
The resulting ink was composed of 5 wt\% Ti nanoparticles, 0.2\,wt\% methylcellulose, 20 \,wt\% ethylene glycol and 20\,wt\% IPA.
Ti PTLs, Ti reference samples and Si reference samples were placed on a heated stage (nominal setpoint temperature 120\,°C) and spray coated with multiple passes in an automated setup.
All samples were dried in air at 65\,°C over night. 
To remove the methylcellulose binder and sinter the nanoparticles, the samples were loaded into a tube furnace, and heated to 380°C under synthetic air, holding the temperature for 30 min.\cite{Simon2017}
To sinter the Ti nanopaticles, the atmosphere was replaced with 5\,\% \ce{H2} in Ar, and the temperature increased to 700°C.\cite{Ernst2024} These conditions were held for 2 h, before letting the furnace cool down over night.

\subsubsection{Catalyst synthesis via ALD}

The ThALD process was performed in a Gemstar 6 reactor (Arradiance) at 220\,°C as discussed in detail in previous studies\cite{Schlicht2018,Schlicht2018a,Haschke2019}.
Briefely, precursor was kept at 90\,°C and exposure consisted of $4\times0.5$\,s micropulses interrupted by a 40\,s soaking step where vacuum pumping of the reactor was interrupted, followed by a 90\,s purge step. 
As counter-reactant, $3\times0.5$\,s micropulses of \ce{O3} were used, followed by a 50\,s purge step.
To monitor the ALD processes, witness samples were prepared by sputter coating a 50 nm Ti thin film onto Si wafers (SiTi reference samples).

The PEALD process was specifically developed for this study, using a Veeco Fiji G2 tool at 220\,°C reactor temperature.
The precursor temperature was set to 115\,°C to stay clear of the maximum temperature of 130\,°C of the precursor manifold, while providing sufficient vapor pressure (Supporting Information Figure~S18b).
The Veeco ALD tool operated with Ar as carrier gas, and under constant vacuum pumping with a timing sequence of 10\,s / 15\,s / 5\,s / 20\,s for Ir precursor exposure / purging / \ce{O2} plasma exposure / purging (see Supporting Information Figure~S17). While the other parameters were chosen based on previous experiences on the same machine\cite{Chen2026}, the Ir precursor exposure time was tuned by analyzing the deposited material on SiTi samples using XPS. 
Owing to the high surface sensitivity of XPS, this allowed obtaining a basic set of parameters with only 50 cycles per ALD run, thereby making good use of the scarce Ir(EtCp)(CHD) precursor.
To determine the saturation pulse time, we calculated the atomic percentage ratios of Ir/(Ir+Ti); beyond 10\,s exposure, the ratio showed no substantial increase, and the exposure time was therefore fixed to 10\,s for all following experiments. 

Up to 300 cycles were deposited by loading the reactor with numerous samples (SiTi reference and PTLs), and consecutively removing samples after the desired number of cycles.
The range of cycles from 300 to 600 samples was covered by depositing another 300 cycles onto selected samples that were previously coated with a variable number of cycles.

\subsubsection{Electrode loading characterization}
The amount of deposited Ir on $5\times10$\,mm$^2$ reference samples from PTL material was determined via instrumental neutron activation analysis (INAA) at the Reactor Institute Delft.
The area-specific loading was calculated while taking into account the coating of substrate edges as described in the Supporting Information.
We note that loading characterization via ICP-MS turned out to be unreliable, as Ir is difficult to digest in conventional analyte solutions.
However, ICP-MS was used to determine the loading of the ALD-made Pt cathode PTEs using a reference sample.\cite{Chen2026}

\subsubsection{XPS analysis}

Surface composition and chemical states were analyzed via XPS using a ThermoFisher Scientific Nexsa G2 instrument. 
Both survey and high-resolution element-specific scans were performed on ALD-coated flat SiTi reference samples.
After charge correction using the C 1s peak at 284.8\,eV, XPS fitting followed the specific procedure for Ir and \ce{IrO2} by Freakly et al.\cite{Freakley2017}
The doublet separations for Ir 4f$_{7/2}$ and 4f$_{5/2}$ for Ir and \ce{IrO2}, respectively, and the peak shift from Ir to \ce{IrO2} were constrained.
The FWHM of the Ir peaks was fitted to the sample with higher metallic Ir content, and fixed to the same value for the sample that mostly showed \ce{IrO2}, and vice versa. For the O 1s fits, the relative positions of the anhydrous oxide, satellite and hydroxide peak were constrained\cite{Freakley2017}.

\subsubsection{Electrochemical characterization}

All anodes were tested on a custom-built PEMWE test bench at atmospheric pressure. 
Serpentine flow fields from Dioxide Materials (Boca Ranton, FL, USA) with 5\,cm$^2$ nominal area were used as endplates, using stainless steel on the cathode, and titanium on the anode side.
The active area was limited to 4\,cm$^2$ by using 2\,cm\,x\,2\,cm electrodes and PTLs, and gaskets with corresponding cut-outs. 
For the anode side, a 1\,mm thick gasket was cut out of PTFE sheet material (Polyfluor, The Netherlands), and used to embed either the ALD-made anode PTEs or, in the case of the CCM anode, a Ti felt as PTL.
On the cathode side, gaskets were cut from 100\,µm thick PTFE sheets to embed the cathode PTEs. The latter were based on 250\,µm thick Freudenberg H23C6, onto which Pt was deposited via direct ALD\cite{Chen2026}.
To ensure a homogeneous pressure distribution, an additional 1\,mm thick Ti felt and PTFE gasket was placed between the cathode flow field and the cathode PTE.
Nafion\texttrademark~117 membranes were pretreated for 1\,h in 1\,M \ce{H2SO4} at 80\,°C, and subsequently for 1\,h in deionized water at 80\,°C, before mounting the cell.
A single-sided commercial CCM reference (QuinTech, Germany) with 1.4\,mg$\cdot$cm$^{-2}$ was treated in the same way.
One ThALD 150c MPL anode (see Figure~\ref{fig:XtraNafion}) was additionally spray coated with a thin Nafion\texttrademark~layer. 40\,µl of Nafion\texttrademark~D-521 dispersion (5\% w/w) was diluted in 1.5\,ml IPA and spray coated onto the anode that was placed onto an 80\, °C surface in multiple passes in an automated spraycoater. 

The PEMWE cell was assembled using eight M5 bolts tightened to 5\,Nm. 
Heating pads (Dioxide Materials) were used on either side of the cell to set a temperature of 80\,°C, and controlled using the thermocouple inserted into the endplate, and a PID controller. 
The anode side was fed with preheated Milli-Q water via a peristaltic pump (Minipuls 3, Gilson, The Netherlands) with a flow rate of 25\,ml$\cdot$min$^{-1}$. 
The anode water was recirculated through a laboratory glass bottle, which served as both reservoir and gas-liquid separator. 
On the cathode side the water inlet was capped off, and the carry-over water collected in another laboratory glass bottle.

A VSP-300 potentiostat (Biologic, France) with a 10\,A booster was used to carry out the characterization sequence in which the cell voltage was limited to 2\,V:
1)~Each MEA was first conditioned at 0.1\Acm for 5\,min to ensure the right oxidation state of the catalysts\cite{JRC_Pols}.
2)~The polarization curves were recorded in galvanostatic mode in 34 current steps\cite{JRC_Pols} between 10$^{-3}$\Acm and 2.5\Acm, and each step was held for 150\,s. The current range of the potentiostat was set to its maximum of 10\,A.
3)~The polarization data for the Tafel plots was obtained similarly, however, setting the potentiostat current range to 1\,A, which allowed more accurate control of currents in the range from 10$^{-3}$\Acm to 10$^{-1}$\Acm with logarithmic spacing as displayed in the Tafel plots throughout the manuscript.
4)~Subsequently, electrochemical impedance spectroscopy was performed in galvanostatic mode at the current densities marked in Figure~\ref{fig:Etareconstruct}.
In order to keep the cell voltage below 2.0\,V and thereby prevent premature cell degradation in ThALD 75c MPL, the full current density range for up to 2.5\Acm for EIS was only tested after the ADT sequence (described below).
The perturbation amplitude was set to 2\% of the DC current, and the frequency was scanned from 100\,kHz to 100\,mHz with six points per decade\cite{JRC_EIS}.
Six measurements were averaged per frequency.
The entire sequence 1)--4) was repeated for the PEALD anodes on bare felts, respectively, as these showed substantial performance improvements during testing.

Selected MEAs were tested regarding durability after replacing the anode water supply with fresh Milli-Q water.
The cells were run in galvanostatic mode at 1\Acm for 24\,h while recording the cell voltage.
Subsequently, the MEAs exposed to an ADT by alternating the cell voltage between 1.45 and 2.0\,V for 25,000 cycles (Figure~\ref{fig:stability}c).
After the ADT, the Milli-Q water was again refreshed, and the MEA run at 1\Acm for at least 3\,h to drive out potentially contaminated water, before repeating characterization sequence 1)--4).

For the \ce{CO2} electrolysis experiment, the anode PTE was mounted in the same way as for the PEMWE experiments, however, circulating 200\,ml of 0.1 M \ce{CsHCO3} as anolyte which was prepared from CsHCO\textsubscript{3} salt (purity 99.9\% purchased from Sigma-Aldrich) and deionized water.
The remainder of the cell assembly followed the procedure previously reported by Montfort et al.\cite{Montfort2023}.
For the cathode, a 100\,nm thick Ag layer was sputter coated onto a carbon-based PTL (SLG Carbon Sigracet\textregistered~39 BB).
The resulting PTE was mounted in the cell using a 250\,µm thick silicone gasket (Laufenberg, Germany).
Anode and cathode were separated by a 40\,µm thick AEM (PiperION\textregistered). The membrane was activated in 1\,M KOH for at least 24\,h prior to usage, and washed in deionized water right before cell assembly. The temperature of the cell environment was regulated by an oven to 25$^{\circ}$C.

\subsubsection{Processing of electrochemical data}
EIS data was fitted using the impedance.py open source package\cite{Murbach2020}, yielding the individual resistance contributions for the model in Figure~\ref{fig:geis}.
The series resistance-free cell voltage was calculated using the $R_{\Omega}$ value obtained from fitting.

Polarization curves were obtained from the galvanostatic measurements by taking the mean of current and voltage, respectively, of each of the 150\,s steps, while discarding the first 30\,s and the last 10\,s.
The low current polarization data shown in Tafel plots were obtained in the same way, and Tafel slopes $b$ and apparent exchange current densities $j_0$ were obtained by fitting the Tafel equation
\begin{equation}\label{eq:Tafel}
   V_\text{cell}(j)-j\cdot R_\Omega - E_\text{rev} = \eta_\text{act,Tafel}= b \log_{10}(\frac{j}{j_0})
\end{equation}
to the data, where $E_\text{rev}=1.168$\,V at ambient pressure and 80\,°C.\cite{BerntIonomer2016}
However, we note that $j_0$ is obtained by extrapolating the current density over several orders of magnitude, leading to high uncertainties.\cite{Matsumoto1986}
Therefore, $j_0$ is not discussed explicitly, but only used to plot $\eta_\text{act,Tafel}$ in Figure~\ref{fig:Etareconstruct}.

\begin{acknowledgement}

We thank Mehmet Sarilar from Delft Reactor Institute for service around the INAA measurements and 
Joost Middelkoop for technical support with PEALD, spray coating and sintering. 
We also thank Agnes Schipper and Lars Bannenberg for fruitful discussions regarding XRR, and Bart Boshuizen for supporting with XPS characterization.

TU Delft gratefully acknowledges financial support from EU project ICONIC (no. 101115204) and ECCM KICkstart DE-NL project, supported by the Dutch Research Council (NWO, no. KICH1.ED04.20.023).

\end{acknowledgement}

\begin{suppinfo}

Electronic Supporting Information is available online.

\end{suppinfo}

\clearpage
\bibliography{dALDIridium}

\end{document}